\documentclass[12pt,a4paper]{article}
\usepackage{times}
\usepackage{a4wide}
\usepackage{amsfonts}
\usepackage{amssymb}
\usepackage{amsmath}
\usepackage{bbm}
\usepackage{booktabs} % for much better looking tables
\usepackage{array}
\usepackage{ifpdf}
\ifpdf
\usepackage[pdftex,unicode,implicit]{hyperref}
\hypersetup{%
  pdftitle    = { New  \texorpdfstring{$E_{7(7)}$}{E77}  Invariants  and  Amplitudes},
  pdfkeywords = {duality, supergravity, finiteness, amplitudes},
  pdfauthor   = {Renata Kallosh, Tomas Ort\'{\i}n},
  plainpages  = true,
  colorlinks  = true,
  citecolor   = blue,
  urlcolor    = red,
  linkcolor   = black
}
\newcommand{\hepth}[1]{{\tt
\href{http://www.arXiv.org/abs/hep-th/#1}{hep-th/#1}}}

\newcommand{\arxiv}[1]{{\tt
\href{http://www.arXiv.org/abs/#1}{arXiv:#1}}}
\else
  \usepackage[dvips]{graphicx}
  \usepackage[unicode,implicit]{hyperref}
  \newcommand{\hepth}[1]{{\tt hep-th/#1}}

  \newcommand{\arxiv}[1]{{\tt arXiv:#1}}
\fi
\makeatletter
\@addtoreset{equation}{section}
\makeatother

\DeclareFontFamily{U}{rsf}{}
\DeclareFontShape{U}{rsf}{m}{n}{
  <5> <6> rsfs5 <7> <8> <9> rsfs7 <10-> rsfs10}{}
\DeclareMathAlphabet\Scr{U}{rsf}{m}{n}
\begin{document}

~\vspace{-2cm}
\begin{flushright}
\small
SU-ITP-12/15\\
IFT-UAM/CSIC-12-43\\
%\texttt{arXiv:YYMM.NNNN [hep-th]}\\
May 18\textsuperscript{th}, 2012\\
\normalsize
\end{flushright}

\vspace{2cm}

\begin{center}

  {\Large\bf  New  $E_{7(7)}$  Invariants  and  Amplitudes}

\vspace{2cm}

\renewcommand{\thefootnote}{\alph{footnote}}
{\sl\large Renata Kallosh$^{\dagger}$}%
\footnote{{\tt kallosh [at] stanford.edu}},
{\sl\large and Tom\'{a}s Ort\'{\i}n$^{\diamond}$}%
\footnote{{\tt Tomas.Ortin [at] csic.es}},
\renewcommand{\thefootnote}{\arabic{footnote}}

\vspace{.5cm}

${}^{\dagger}${\it Stanford Institute for Theoretical Physics and Department of Physics,\\
  Stanford University,  Stanford, CA 94305-4060, USA}\\

\vspace{.4cm}

${}^{\diamond}${\it Instituto de F\'{\i}sica Te\'orica UAM/CSIC\\
C/ Nicol\'as Cabrera, 13--15, 28049 Madrid, Spain}\\

\vspace{2cm}

%%%%%%%%%%%%%%%%%%%%%%%%%%%%%%%%%%%%%%%%%%%%%%%%%%%%%%%%%%%%%%%%%%%%%%

{\bf Abstract}

\end{center}

\begin{quotation}
  \small We construct a new class of manifest $E_{7(7)}$\, duality invariants,
  which generalize the Cartan quartic invariant, familiar from studies of the
  black hole entropy. The new ones, being four-linear, are designed for
  studies of the four-vector amplitudes and to be used, upon
  supersymmetrization, as initial sources of deformation for the non-linear
  higher-derivative generalizations of the linear twisted self-duality
  condition.  We show, however, that the new invariants are inconsistent with
  the expected UV divergent amplitudes in extended supergravities with
  non-degenerate duality groups of type $E7$. When $E_{7(7)}$ degenerates into
  $U(1)$ the new invariants reproduce the recently discovered source of
  deformation for the Born-Infeld duality invariant model with higher
  derivatives relevant to UV divergences of the D3 brane action.  These facts
  may explain the UV properties of perturbative supergravity.
\end{quotation}

\newpage
%%%%%%%%%%%%%%%%%%%%%%%%%%%%%%%%%%%%%%%%%%%%%%%%%%%%%%%%%%%%%%%%%%%%%%
%%%%%%%%%%%%%%%%%%%%%%%%%%%%%%%%%%%%%%%%%%%%%%%%%%%%%%%%%%%%%%%%%%%%%%
%%%%%%%%%%%%%%%%%%%%%%%%%%%%%%%%%%%%%%%%%%%%%%%%%%%%%%%%%%%%%%%%%%%%%%
%%%%%%%%%%%%%%%%%%%%%%%%%%%%%%%%%%%%%%%%%%%%%%%%%%%%%%%%%%%%%%%%%%%%%%
\pagestyle{plain}
%%%%%%%%%%%%%%%%%%%%%%%%%%%%%%%%%%%%%%%%%%%%%%%%%%%%%%%%%%%%%%%%%%%%%%
%%%%%%%%%%%%%%%%%%%%%%%%%%%%%%%%%%%%%%%%%%%%%%%%%%%%%%%%%%%%%%%%%%%%%%
%%%%%%%%%%%%%%%%%%%%%%%%%%%%%%%%%%%%%%%%%%%%%%%%%%%%%%%%%%%%%%%%%%%%%%
%%%%%%%%%%%%%%%%%%%%%%%%%%%%%%%%%%%%%%%%%%%%%%%%%%%%%%%%%%%%%%%%%%%%%%
%%%%%%%%%%%%%%%%%%%%%%%%%%%%%%%%%%%%%%%%%%%%%%%%%%%%%%%%%%%%%%%%%%%%%%

\tableofcontents

%%%%%%%%%%%%%%%%%%%%%%%%%%%%%%%%%%%%%%%%%%%%%%%%%%%%%%%%%%%%%%%%%%%%%%
%%%%%%%%%%%%%%%%%%%%%%%%%%%%%%%%%%%%%%%%%%%%%%%%%%%%%%%%%%%%%%%%%%%%%%
%%%%%%%%%%%%%%%%%%%%%%%%%%%%%%%%%%%%%%%%%%%%%%%%%%%%%%%%%%%%%%%%%%%%%%
%%%%%%%%%%%%%%%%%%%%%%%%%%%%%%%%%%%%%%%%%%%%%%%%%%%%%%%%%%%%%%%%%%%%%%
\newpage

\section*{Introduction}

The new wave of the interest in extended supergravity is due to a
cancellation, unexpected by most supergravity experts, of the 3-loop and
4-loop UV divergences in $\mathcal{N}=8$ \cite{Bern:2007hh,Bern:2008pv} and
3-loop in $\mathcal{N}=4$ \cite{Bern:2012cd,Tourkine:2012ip}.  $\mathcal{N}=8$
supergravity has an $E_{7(7)}$ duality symmetry discovered by Cremmer and
Julia \cite{Cremmer:1979up,de Wit:1982ig}. The dualities of the
$\mathcal{N}>2$ supergravities were first studied by Gaillard and Zumino from
a general point of view in \cite{Gaillard:1981rj} where it was shown that
these symmetries require the conservation of the Noether-Gaillard-Zumino (NGZ)
current and, equivalently, the fulfillment of the NGZ identities. These
identities were studied with respect to $\mathcal{N}=8$ supergravity in
\cite{Kallosh:2011dp}, where it was shown that adding the higher loop
counterterms to the action would break the $E_{7(7)}$\, NGZ current
conservation. It was, however, suggested by Bossard and Nicolai
\cite{Bossard:2011ij} that the problem may be cured by adding some extra
terms, in addition to the counterterm, so that the NGZ current conservation is
restored.
  
The original proposal to deform the linear twisted self-duality condition made
in \cite{Bossard:2011ij} was studied in \cite{Carrasco:2011jv} and was shown
to be incomplete: it required a significant generalization even to derive the
familiar Born-Infeld model by this method. A generalized procedure for
deriving new models with duality symmetry was proposed and developed in
\cite{Carrasco:2011jv}. It turned out to be useful to construct new
duality-invariant models such as a Born-Infeld-type model with higher
derivatives \cite{Chemissany:2011yv} and new $\mathcal{N}=2$ $U(1)$ gauge
theories \cite{Broedel:2012gf}.

To apply the generalized procedure of deformation of the linear twisted
self-duality condition \cite{Bossard:2011ij,Carrasco:2011jv} to
$\mathcal{N}=8$ supergravity it is necessary to find an initial source of
deformation which 

\begin{enumerate}
\item is manifestly $E_{7(7)}$-invariant. This requires the simultaneous use
  of the 28 vector fields present in the action and the 28 dual vector fields,
  associated with the derivative of the action over the vectors, combined in
  doublets which transform in the fundamental (the $\mathbf{56}$) of
  $E_{7(7)}$.
\item constrained by the linear twisted self-duality condition, should match
  the candidate UV divergence. In $\mathcal{N}=8$ supergravity we should
  compare it with the corresponding 4-vector amplitude
  \cite{Kallosh:2008ru,Freedman:2011uc}.
\end{enumerate}

While the construction of initial sources of deformation with these properties
is possible at a bosonic level, we will show that such invariants cannot exist
in the geometric $\mathcal{N}=8$ superspace construction of
\cite{Brink:1979nt,Howe:1981gz}: relaxing the linear twisted self-duality
condition violates the superspace Bianchi identities because the doubling of
vector field degrees of freedom (from 28 to 56) is not possible in the
framework of the superspace, where there is only one superfield and its
derivatives \cite{Brink:1979nt,Howe:1981gz}. A more general superspace or
component construction which may accommodate this doubling of vectors still
has to be produced.

We leave this as an open issue and proceed to do what can be done with the
available techniques, namely to construct new bosonic $E_{7(7)}$\, invariants
from fundamental vector field doublets. Notice that $E_{7(7)}$\, invariants
constructed only of fundamental vector field doublets are rare!  The one we
need, actually, had not yet been constructed and we will do it here.

If we were able to find the proper bosonic $E_{7(7)}$\, invariants, we would
have to supersymmetrize them. But, as we will show, the obstruction for the
amplitudes will show up already at the bosonic level, so there will be no need
to look for the supersymmetric version of the new invariants, at least with
regard to the UV divergence analysis.

The \textit{groups of type} $E7$ \cite{Brown} are groups of linear
transformations leaving invariant two multilinear forms: a skew-symmetric
bilinear and a symmetric four-linear form\footnote{See
  \cite{Ferrara:2011dz,Kallosh:2012ei} for recent applications of this concept
  in Physics.}. $E_{7(7)}$ is the prime example of these groups. By
definition, two invariants of the groups of type $E7$ can always be
constructed using only objects in the fundamental representation
(\textit{fundamentals}): the bilinear symplectic, antisymmetric, and the
four-linear one, which is symmetric. When all the fundamentals are identical,
the would-be quadratic invariant vanishes due to antisymmetry and the
four-linear invariant gives a quartic invariant.

The familiar Cartan-Cremmer-Julia quartic invariant
\cite{CARTAN,Cremmer:1979up} of $E_{7(7)}$ which plays a significant role in
describing $\mathcal{N}=8$ black-hole entropy and its relation to quantum
entanglement of qubits\footnote{See, for example, \cite{Kallosh:1996uy}
  -\cite{Bianchi:2009wj}.}  is a particular example of this mechanism. It can
be obtained from the four-linear invariant of $E_{7(7)}$, whose structure is
based on exceptional Jordan algebra $J_{3}^{\mathbb{O}_{S}}$ over the split
octonions $\mathbb{O}_{S}$ \cite{Gunaydin:2000xr,Gunaydin:2009zza}.  This
four-linear invariant of $E_{7(7)}$ will be used in what follows to construct
new $E_{7(7)}$ invariants relevant to amplitudes and counterterms.  When the
duality group reduces to $U(1)$ these new invariants will reduce to the ones
identified in the recently constructed Born-Infeld models with higher
derivatives \cite{Chemissany:2011yv}. We will then compare these new
$E_{7(7)}$\, invariants with the UV counterterms of $\mathcal{N}=8$
supergravity \cite{Kallosh:1980fi}.

In particular, we will compare the answer with the expression for the 4-point
vector amplitude in $\mathcal{N}=8$ supergravity. Since it is the MHV
amplitude, it predicts the structure of the UV divergences in the 4-vector
sector, starting from the 3-loop level, as shown in \cite{Kallosh:2008ru}.
The analogous explicit contribution to the 4-vector 3-loop UV divergence was
obtained in \cite{Freedman:2011uc}.  All we have to do is to take our new
manifest $E_{7(7)}$\, invariant defined as a functional of the 4 fundamentals
$\mathbf{56}$ (at each of the 4 momenta $p_{I}$), compute its value when
imposing the linear twisted self-duality constraint and compare with the
4-point MHV vector amplitude. We will find that, as long as we consider a
non-degenerate case of groups of type $E7$, the invariants disagree with the
expected UV divergences of the 4-vector amplitudes. Meanwhile, for the case of
degenerate groups of type $E7$, the invariants may agree with the expected UV
divergences of the 4-vector amplitudes. Specifically, in the $U(1)$ duality
case our reduced invariants do reproduce the first loop UV divergence of the
D3 brane action computed in \cite{Shmakova:1999ai} and the 1-loop UV
divergence of the $\mathcal{N}=2$ supersymmetric Born-Infeld action computed
in \cite{DeGiovanni:1999hr}.

Various aspects of $E_{7(7)}$\, symmetry with regard to perturbative
$\mathcal{N}=8$ supergravity were studied before \cite{Kallosh:2008ic}
-\cite{Beisert:2010jx} and some of these results will be used here.

This paper is organized as follows: in Section~\ref{sec-strategy} we introduce
the graviphoton field strength of $\mathcal{N}=8$ supergravity, $T_{IJ\,
  \mu\nu}$ and we discuss the possible construction of manifestly
duality-invariant initial sources of deformation of the linear, twisted,
selfduality constraint using $T_{IJ\, \mu\nu}$ or using the vector field
strengths in the fundamental of $E_{7(7)}$. We argue that with the current
superspace techniques (the argument is explained in detail in
Appendix~\ref{ap-superspace}), such a construction is only possible using the
latter. Section~\ref{sec-invariantsandbhs} is devoted to the general
construction of $E_{7(7)}$ invariants of the electric and magnetic charges (up
to fourth order in charges), which also fill a fundamental representation of
that group. We study the cases of a single charge and of four different
charges and discuss the uses of these invariants in (extremal) black-hole
physics. This section can actually be understood as a preparation for
Section~\ref{sec-invariantsandamplitudes} in which, using the $E_{7(7)}$
invariants just constructed we construct new candidates to $E_{7(7)}$- and
Lorentz-invariant initial sources of deformation and scattering amplitudes,
showing that those computed in the literature disagree with them. In
Section~\ref{sec-reduction} we study the reduction of the invariants to
smaller duality groups. Finally, we discus our results in
Section~\ref{sec-discussion}. The Appendices~\ref{app-t8} and C contain
complementary information on the Green-Schwarz $t^{(8)}$ tensor and on the
split octonions, respectively.

%%%%%%%%%%%%%%%%%%%%%%%%%%%%%%%%%%%%%%%%%%%%%%%%%%%%%%%%%%%%%%%%%%%%%%
%%%%%%%%%%%%%%%%%%%%%%%%%%%%%%%%%%%%%%%%%%%%%%%%%%%%%%%%%%%%%%%%%%%%%%
%%%%%%%%%%%%%%%%%%%%%%%%%%%%%%%%%%%%%%%%%%%%%%%%%%%%%%%%%%%%%%%%%%%%%%
%%%%%%%%%%%%%%%%%%%%%%%%%%%%%%%%%%%%%%%%%%%%%%%%%%%%%%%%%%%%%%%%%%%%%%
%%%%%%%%%%%%%%%%%%%%%%%%%%%%%%%%%%%%%%%%%%%%%%%%%%%%%%%%%%%%%%%%%%%%%%

\section{The strategy: graviphoton, scalars and vectors}
\label{sec-strategy}

%%%%%%%%%%%%%%%%%%%%%%%%%%%%%%%%%%%%%%%%%%%%%%%%%%%%%%%%%%%%%%%%%%%%%%
%%%%%%%%%%%%%%%%%%%%%%%%%%%%%%%%%%%%%%%%%%%%%%%%%%%%%%%%%%%%%%%%%%%%%%
%%%%%%%%%%%%%%%%%%%%%%%%%%%%%%%%%%%%%%%%%%%%%%%%%%%%%%%%%%%%%%%%%%%%%%
%%%%%%%%%%%%%%%%%%%%%%%%%%%%%%%%%%%%%%%%%%%%%%%%%%%%%%%%%%%%%%%%%%%%%%
%%%%%%%%%%%%%%%%%%%%%%%%%%%%%%%%%%%%%%%%%%%%%%%%%%%%%%%%%%%%%%%%%%%%%%

\subsection{The graviphotons of  $\mathcal{N}=8$ supergravity}
\label{sec-graviphotons}

The relevant sector of the Lagrangian of $\mathcal{N}=8,d=4$ classical 
supergravity\footnote{Here we follow the tested conventions of
  Ref.~\cite{Meessen:2010fh} essentially taken from
  Ref.~\cite{Andrianopoli:1996ve} and introduce the notation and the main
  definitions we are going to need.}  is
\begin{equation}
\label{eq:Lagrangian}
L= 2\Im{\rm m}\, \mathcal{N}_{\Lambda\Sigma}F^{\Lambda}{}_{\mu\nu}   
F^{\Sigma\, \mu\nu}
-2\Re{\rm e}\, \mathcal{N}_{\Lambda\Sigma}F^{\Lambda}{}_{\mu\nu}   
\tilde{F}^{\Sigma\, \mu\nu}\, ,
\end{equation}
where $\tilde{F}$ is the Hodge dual of $F$, defined in
Eq.~(\ref{eq:Hodgedualdef}), $\mathcal{N}_{\Lambda\Sigma}$ is the
scalar-dependent period matrix (whose definig property is given in
Eq.~(\ref{eq:periodmatrixdef}) below) and where the indices
$\Lambda,\Sigma=1,\cdots,28$. Later on, each of these indices will be replaced
by an antisymmetrized pair of indices $i,j=1,\cdots,8$.

The electromagnetic dual of $F^{\Lambda}$ (\textit{magnetic} field strengths)
$G_{\Lambda}$ are defined by 
\begin{equation}
\label{eq:magneticdef}
\tilde{G}_{\Lambda\, \mu\nu}
\equiv 
-\tfrac{1}{4}\frac{\partial L}{\partial F^{\Lambda\, \mu\nu}}
=
-\Im{\rm m}\, \mathcal{N}_{\Lambda\Sigma}
F^{\Sigma}{}_{\mu\nu}
+\Re{\rm e}\, \mathcal{N}_{\Lambda\Sigma}
\tilde{F}^{\Sigma}{}_{\mu\nu}\, ,
\end{equation}
or, equivalently,
\begin{equation}
\label{eq:constraint}
G_{\Lambda}{}^{+}=\overline{\mathcal{N}}_{\Lambda\Sigma}F^{\Sigma\, +}\, .  
\end{equation}

In most of what follows we will consider $F^{\Lambda}$ and $G_{\Lambda}$ as
independent variables. In particular, this will mean that the $G_{\Lambda}$
are independent of the scalars and Eq.~(\ref{eq:constraint}) is not
satisfied. The dependence will be reintroduced only after imposing the
constraint Eq.~(\ref{eq:constraint}), known as \textit{linear twisted
  self-duality constraint}. Observe that, given the definition of the magnetic
field strengths Eq.~(\ref{eq:magneticdef}) this constraint contains
information enough to reconstruct the Lagrangian Eq.~(\ref{eq:Lagrangian}).

With these field strengths we can construct a $56$-dimensional real symplectic
vector of field strengths
\begin{equation}
\mathcal{F} \equiv 
\left(
  \begin{array}{c}
  F^{\Lambda} \\ G_{\Lambda} \\  
  \end{array}
\right)\, .
\end{equation}
that transforms in the $\mathbf{56}$ of $E_{7(7)}\subset Sp(56,\mathbb{R}))$.

The scalars of the theory are described by the symplectic section
\begin{equation}
\label{eq:symplecticsection}
\mathcal{V}_{IJ}
\equiv
\left(
  \begin{array}{c}
  f^{\Lambda}{}_{IJ} \\ h_{\Lambda\, IJ} \\  
  \end{array}
\right)\, ,
\end{equation}
where $I,J=1,\cdots,8$ are an antisymmetric pair of indices that are raised
and lowered by complex conjugation. The period matrix is defined by the
property
\begin{equation}
\label{eq:periodmatrixdef}
h_{\Lambda\, IJ} = \mathcal{N}_{\Lambda\Sigma}f^{\Sigma}{}_{IJ}\, .  
\end{equation}

The components of the section $\mathcal{V}_{IJ}$ are related to the components
of the coset representative 
\begin{equation}
{\Scr V} \equiv 
\left(
  \begin{array}{cc}
A & B \\
C & D\\ 
  \end{array}
\right)\,\,   \in E_{7(7)}/SU(8) \subset
Sp(56,\mathbb{R})\, ,\,\,\,\,
\Rightarrow 
\left\{
  \begin{array}{rcl}
A^{T}C-C^{T}A & = & 0\, ,\\
B^{T}D-D^{T}B & = & 0\, ,\\
A^{T}D-C^{T}B & = & \mathbbm{1}_{28\times 28}\, ,\\
\end{array}
\right.
\end{equation}
by 
\begin{equation}
f = \tfrac{1}{\sqrt{2}}(A-iB)\, ,
\hspace{1cm}
h = \tfrac{1}{\sqrt{2}}(C-iD)\, .  
\end{equation}
This relation of the components of the section $\mathcal{V}_{IJ}$ with the
components of the symplectic $E_{7(7)}/SU(8)$ coset representative imply the
constraints\footnote{For the symplectic product $\langle ~ \mid ~ \rangle$, we
  use the convention
\begin{equation}
\langle  \mathcal{A}\mid \mathcal{B}\rangle \equiv
  \mathcal{B}^{\Lambda}\mathcal{A}_{\Lambda} 
-\mathcal{B}_{\Lambda}\mathcal{A}^{\Lambda}\, .
\end{equation}
}
\begin{equation}
\langle \mathcal{V}_{IJ}\mid\overline{\mathcal{V}}^{\, KL}\rangle 
 =    
-2i\delta_{IJ}{}^{KL}\, , 
\hspace{1cm}
\langle \mathcal{V}_{IJ}\mid\mathcal{V}_{KL}\rangle =0\, .
\end{equation}
It also implies that, when we perform a global $E_{7(7)}$ transformations that
acts linearly on the $\Lambda, \Sigma$ indices, we have to act linearly with a
$SU(8)$ compensating transformation (sometimes called ``local'' because they
depend on the scalar fields) on the indices $I,J,\ldots$. Thus, the indices
$I,J,\ldots$ are referred to as $SU(8)$ indices. There is another, completely
different kind of $SU(8)$ indices that will enter the game later on and that
will be denoted by indices $A,B,C,\ldots$.

The graviphoton field strength is defined by
\begin{equation}
\label{eq:graviphotondef}
T_{IJ}  
\equiv
\langle \mathcal{V}_{IJ}\mid \mathcal{F} \rangle\, , 
\end{equation}
and its self- and anti-selfdual parts, defined in Eq.~(\ref{eq:selfdualdef})
are
\begin{equation}
T_{IJ}{}^{\pm}  
\equiv
\langle \mathcal{V}_{IJ}\mid \mathcal{F}{}^{\pm}   \rangle\, . 
\end{equation}
They all transform under compensating $SU(8)$ transformations only.  Since the
$SU(8)$ tensor $T_{IJ}$ is complex, we have
\begin{equation}
T^{IJ\, \pm} = \overline{(T_{IJ}{}^{\mp})}\, .  
\end{equation}
Finally, the linear twisted self-duality constraint Eq.~(\ref{eq:constraint}),
is equivalent to the vanishing of some of these objects:
\begin{equation}
\label{linear}
\overline{T}^{IJ\, +} = \overline{(T_{IJ}{}^{-})} =0\, . 
\end{equation}

This vanishing has important implications. Using the Maurer-Cartan equations 
\begin{equation}
\label{eq:Maurer-Cartan}
\mathfrak{D}\mathcal{V}_{IJ} =
\tfrac{1}{2}\mathcal{P}_{IJKL}\overline{\mathcal{V}}^{\, KL}\, ,  
\end{equation}
where $\mathfrak{D}$ is the $SU(8)$-covariant derivative and
$\mathcal{P}_{IJKL}$ the Vielbein 1-form on the scalar manifold, and the
definition of the graviphoton field strength (\ref{eq:graviphotondef}) we find
\begin{equation}
\mathfrak{D}T_{IJ} =
\tfrac{1}{2}\mathcal{P}_{IJKL}\wedge \overline{T}^{\, KL}\, ,  
\end{equation}
and its complex conjugate.

%%%%%%%%%%%%%%%%%%%%%%%%%%%%%%%%%%%%%%%%%%%%%%%%%%%%%%%%%%%%%%%%%%%%%%
%%%%%%%%%%%%%%%%%%%%%%%%%%%%%%%%%%%%%%%%%%%%%%%%%%%%%%%%%%%%%%%%%%%%%%
%%%%%%%%%%%%%%%%%%%%%%%%%%%%%%%%%%%%%%%%%%%%%%%%%%%%%%%%%%%%%%%%%%%%%%
%%%%%%%%%%%%%%%%%%%%%%%%%%%%%%%%%%%%%%%%%%%%%%%%%%%%%%%%%%%%%%%%%%%%%%
%%%%%%%%%%%%%%%%%%%%%%%%%%%%%%%%%%%%%%%%%%%%%%%%%%%%%%%%%%%%%%%%%%%%%%

\subsubsection{Manifestly $E_{7(7)}$-invariant terms from graviphotons}
\label{sec-invariantsfromgraviphotons}

Since, as we have stressed above, the graviphoton field strength $T_{IJ}$
(\ref{eq:graviphotondef}) only transforms under the induced $SU(8)$
compensating transformations which act linearly on the indices $I,J,\ldots$,
it is trivial to construct terms which are manifestly invariant under the full
$E_{7(7)}$ group by contracting all the $SU(8)$ indices in the standard way.
On the other hand, in classical $\mathcal{N}=8$ supergravity the linear
twisted self-duality constraint Eq.~(\ref{linear}) is valid by construction
and the only available non-vanishing tensors that we can use are $T^{IJ\, -}$
and its complex conjugate $T_{IJ}{}^{+}$.

In the context of the deformation procedure
\cite{Bossard:2011ij,Carrasco:2011jv,Chemissany:2011yv,Broedel:2012gf} the
linear twisted self-duality condition Eq.~(\ref{linear}) has to be deformed to
accommodate the counterterms. A priory, one may consider two possibilities for
the initial source of deformation $\mathcal{I}$ from which the counterterms
are derived by an interative procedure:

\begin{enumerate}
\item ${\cal I}(F, G; \phi)= {\cal I}(T)$ depends on the graviphoton and its
  $SU(8)$-covariant derivatives. In particular, this is a property of the
  counterterms constructed in \cite{Kallosh:1980fi}. The initial source of
  deformation ${\cal I}$ must be a manifestly Lorentz- and
  $E_{7(7)}$-invariant expression depending on the fundamental vector doublet
  $\mathcal{F}=(F, G)$ via the graviphoton components $T^{IJ\, -},
  T_{IJ}{}^{+}$ which do not vanish at the linear level. 

  Given that initial source of deformation one defines the \textit{deformed
    twisted self-duality constraint}
  \begin{equation}
    T^{IJ\, +}{}_{\mu\nu} = \frac{\delta {\cal I} (T)}{\delta T_{IJ}{}^{+\,
        \mu\nu}}\, ,
  \end{equation} 
  which can be solved by iteration. At the lowest (linear) order the r.h.s.~of
  this expression vanishes and one recovers the linear twisted self-duality
  constraint Eq.~(\ref{linear}). At higher orders the r.h.s.~produces
  non-linear deformations of that constraint. Just as the linear constraint
  contains enough information to resconstruct the Lagrangian
  Eq.~(\ref{eq:Lagrangian}), the non-linear deformations lead to Lagrangians
  with terms of higher orders in the graviphoton field strength.

\item ${\cal I}(F, G)$ depends only on the $E_{7(7)}$\, fundamental vector
  field strength doublet $\mathcal{F}=(F, G)$. In such case, the deformation
  procedure may be described as the duality covariant variation of ${\cal
    I}(F, G)$ over the doublet, using the scalar-dependent metric, as proposed
  in \cite{Bossard:2011ij}.
\end{enumerate}

Our first step is to prove that the first possibility encounters an
obstruction. We will find that the supersymmetric version of the linear
twisted self-duality condition (\ref{linear}) is required for the Bianchi
identities of the superspace \cite{Brink:1979nt,Howe:1981gz} where the
invariants are constructed. In particular, we will focus on the superspace
solution of Bianchi identities\footnote{The $SU(8)$ indices that we denote by
  $I,J,\ldots$ in this paper are denoted by $i,j,\ldots$ in
  \cite{Brink:1979nt,Howe:1981gz}.}
\begin{equation}
F_{\alpha \dot{\alpha}, \beta \dot{\beta}, IJ}(x, \theta) 
= 
-i \epsilon_{\dot{\alpha} \dot{\beta}} 
M_{\alpha \beta IJ}(x, \theta)
-i \epsilon_{ \alpha \beta} 
\overline{N}_{\dot{\alpha} \dot{\beta} IJ}(x, \theta)\, , 
\end{equation} 
which allows, in principle to have a 56-component vector doublet $(M_{\alpha
  \beta IJ}, N_{\alpha \beta}{}^{IJ})$ and the conjugate one
$(\overline{M}_{\dot \alpha \dot{\beta}}^{IJ}, \overline{N}_{\dot{\alpha}
  \dot{\beta} IJ})$, since where there are 28 vectors in $M$ and 28 in
$N$. However, the superspace integrability condition for the existence of the
supervielbein that relates $E_{7(7)}$\, to $SU(8)$ in the form
\begin{equation}
\label{curv}
R^{I}{}_{J}(x, \theta)= -
\tfrac{1}{3} \overline{\cal P}^{IKLM}(x, \theta) \wedge \mathcal{P}_{JKLM}(x, \theta)\, ,
\end{equation}
requires that the second part of the vector multiplet, $\overline{N}_{\dot{\alpha}
  \dot{\beta} IJ}, \overline{N}_{\dot{\alpha} \dot{\beta} IJ}$ is constrained to be a
bilinear of the fermion superfields and cannot be independent 
\begin{equation}
\label{N}
N_{\alpha\beta}^{IJ}(x, \theta)
= 
-\tfrac{1}{72} \epsilon^{IJKLMPQR}\chi_{KLM\, \alpha}
 \chi_{PQR\, \beta} (x, \theta)\, .
\end{equation}
Therefore, the existence of the scalar field-dependent sections of an $Sp(56,
\mathbb{R})$ bundle over the $E_{7(7)}/SU(8)$ coset space $h_{\Lambda\,
  IJ}(\phi) $ and $f^{\Lambda}{}_{IJ}(\phi)$ would be inconsistent with
supersymmetry if the vectors were doubled, which requires the superfield
$\overline{N}_{\dot{\alpha} \dot{\beta} IJ}$
to be independent. We provide the derivation of this assertion in Appendix A.

The observation above precludes the deformation procedure
\cite{Bossard:2011ij,Carrasco:2011jv,Chemissany:2011yv,Broedel:2012gf} for
making consistent $\mathcal{N}=8$ supergravity\footnote{At least with means
  available. New constructions may change the situation in the future, if
  successful.}  with counterterms, since the cornerstone of this procedure is
the source of deformation depending on unconstrained doublet of
$E_{7(7)}$. There is no such scalar-dependent graviphoton superfield doublet
in superspace \cite{Brink:1979nt,Howe:1981gz} and the alternatives are also
not available, at present.

%%%%%%%%%%%%%%%%%%%%%%%%%%%%%%%%%%%%%%%%%%%%%%%%%%%%%%%%%%%%%%%%%%%%%%
%%%%%%%%%%%%%%%%%%%%%%%%%%%%%%%%%%%%%%%%%%%%%%%%%%%%%%%%%%%%%%%%%%%%%%
%%%%%%%%%%%%%%%%%%%%%%%%%%%%%%%%%%%%%%%%%%%%%%%%%%%%%%%%%%%%%%%%%%%%%%
%%%%%%%%%%%%%%%%%%%%%%%%%%%%%%%%%%%%%%%%%%%%%%%%%%%%%%%%%%%%%%%%%%%%%%
%%%%%%%%%%%%%%%%%%%%%%%%%%%%%%%%%%%%%%%%%%%%%%%%%%%%%%%%%%%%%%%%%%%%%%
 
\subsection{$E_{7(7)}$\, invariants from fundamentals}
\label{sec-invariantsfromfundamentals}

Here we will construct the $E_{7(7)}$\, invariants relevant for the amplitudes
in $\mathcal{N}=8$ supergravity.  Since $E_{7(7)}$\, acts on the doublet
linearly and there are no scalars with their shift symmetry, we are just
looking at the 4-vector local invariants/amplitudes.  The two major
ingredients in our construction are:

\begin{enumerate}
\item The G\"unaydin-Koepsell-Nicolai construction of the four-linear
  invariant of $E_{7(7)}$\, based on exceptional Jordan algebra $J_{3}
  ^{\mathbb{O}_S }$ over the split octonions $\mathbb{O}_S$
  \cite{Gunaydin:2000xr}. The symplectic invariant and the triple Jordan
  product will be used to produce an eighth-rank Lorentz tensor
  \begin{equation}
    J_{(8)}(\mathcal{F}_{1\, \mu_{1}\nu_{1}}, \mathcal{F}_{2\, \mu_{2}\nu_{2}}, 
\mathcal{F}_{3\, \mu_{3}\nu_{3}}, \mathcal{F}_{4\, \mu_{4}\nu_{4}})\, , 
  \end{equation}
  antisymmetric in each pair of indices $\mu_{I}\nu_{I}$, with $I=1,2,3,4$,
  and symmetric under the exchange of such pairs. Each of the four entries is
  a fundamental $\mathbf{56}$ taken at one of the four momenta,
  $\mathcal{F}_{I\, \mu\nu}\equiv \Big ( F^{ij}{}_{\mu\nu} (p_{I}) , G_{ij\,
    \mu\nu} (p_{I}) \Big )$ so $J_{(8)}$ must be associated to the 4-vector
  amplitude depending on 4 momenta $p_{I}$.

\item The Green-Schwarz eighth-rank tensor $t^{(8)}$ (a symmetrized sum of
  products of four Kronecker $\delta$'s \cite{Green:1981xx} which provides the
  kinematic factor of the open string tree-level 4-point amplitude and has the
  same index symmetry as $J_{(8)}$). We will use it here to construct an
  $E_{7(7)}$- and Lorentz-invariant object from the $E_{7(7)}$-invariant
  tensor $J_{(8)}$ by contrating both tensors. We will also introduce some
  function of Mandelstam variables, so our candidate for initial source of
  deformation takes the general form\footnote{When we use the complex basis,
    that will be introduced in the next sectoin, it will take the form
  \begin{equation}
    t^{(8)}\cdot \diamondsuit_{(8)}\,  f(s,t,u)\, .
  \end{equation}
}
  \begin{equation}
    t^{(8)}\cdot J_{(8)}\,  f(s,t,u)\, .
  \end{equation}
\end{enumerate}

We will compare this proposal with the expression for the 4-point vector
amplitude in $\mathcal{N}=8$ supergravity. Since it is the MHV amplitude, it
predicts the structure of the UV divergences in the 4-vector sector, starting
from the 3-loop level, as shown in \cite{Kallosh:2008ru}.  The analogous
explicit contribution to the 4-vector 3-loop UV divergence was obtained in
\cite{Freedman:2011uc}.  All we have to do is to take the new manifest
$E_{7(7)}$\, invariant defined as a functional of the 4 unconstrained
fundamentals $\mathbf{56}$ (at each of the 4 momenta $p_{I}$) given above,
compute its value when imposing the linear twisted self-duality constraint,
and compare with the 4-point MHV vector amplitude.

In this comparison we will find a discrepancy. As will review in the next
section, the four-linear invariant in the complex ($SU(8)$) basis consists of
3 terms.  Each of them is manifestly invariant under the $SU(8)$ subgroup
of$E_{7(7)}$ which acts diagonally in the complex basis. However, the
invariance under the off-diagonal part of $E_{7(7)}$\, is only achieved when
those 3 terms are combined with very specific coefficients. In the simplest
case, when there is only one $\mathbf{56}$, $ (\mathcal{F}_{AB},
\overline{\mathcal{F}}^{AB})$, the four-linear invariant reduces to the
Cartan-Cremmer-Julia \cite{Cremmer:1979up} quartic invariant\footnote{We
  ignore here the Lorentz indices of the vector field strengths, which do not
  play any role in this discussion.}
\begin{equation}
\label{diamond}
  \diamondsuit  (\mathcal{F})
  = 
  \mathrm{Tr}_{SU(8)} (\mathcal{F}\overline{\mathcal{F}}\mathcal{F}\overline{\mathcal{F}}) 
  -\tfrac{1}{4} \left[\mathrm{Tr}_{SU(8)} (\mathcal{F}\overline{\mathcal{F}}) \right]^{2} 
  +\tfrac{1}{4}\mathrm{Pf}_{SU(8)}\, ||\mathcal{F}||
  +\tfrac{1}{4}\mathrm{Pf}_{SU(8)}\, ||\overline{\mathcal{F}}||\, ,
\end{equation}
and the specific coefficients are $-1/4, 1/4, 1/4$. In our case, each of the 4
factors $\mathcal{F}$ is taken at different momentum and we form a Lorentz
scalar as a product of two Lorentz eight-tensors, but the structure is
analogous to (\ref{diamond}): we have $SU(8)$ traces of 4 operators, squares f
the traces of 2 operators and  Pfaffian.

When we will construct our new manifest $E_{7(7)}$\, invariant designed for
the initial source of deformation of the linear twisted self-duality
condition, we will first test it when the linear twisted self-duality
condition is imposed. We will find that the Pfaffians vanish but the $
\mathrm{Tr}_{SU(8)}
(\mathcal{F}\overline{\mathcal{F}}\mathcal{F}\overline{\mathcal{F}}) $ and
$\left[\mathrm{Tr}_{SU(8)} (\mathcal{F}\overline{\mathcal{F}}) \right]^{2}$
terms will appear with a relative factor which is not just $-1/4$, but a
non-trivial function of the Mandelstam variables ($\frac{st+ su}{tu}$). This
factor breaks $E_{7(7)}$\, symmetry. This also implies that the counterterms
are $E_{7(7)}$-invariant only when the linear twisted self-duality condition
is valid.

Meanwhile, in case of $U(1)$ duality, one finds the initial source of the
deformation manifestly $U(1)$ invariant in all cases studied in
\cite{Carrasco:2011jv,Chemissany:2011yv,Broedel:2012gf}. One finds that this
symmetry takes place independently as to whether the linear twisted
self-duality constraint is imposed or not. This made it possible to develop
the proposal of \cite{Bossard:2011ij} for all $U(1)$ models and it was
possible to produce novel models with a consistent NGZ $U(1)$ current
conservation, which was done in
\cite{Carrasco:2011jv,Chemissany:2011yv,Broedel:2012gf}.

Below we will show that the UV divergences in $\mathcal{N}=8$ supergravity
starting from the 3-loop level would break the $E_{7(7)}$\, current
conservation.  We will show that the procedure proposed in
\cite{Bossard:2011ij}, as different from the $U(1)$ models, cannot be
improved. It is inconsistent with the structure of the four-linear invariant
of $E_{7(7)}$\, based on exceptional Jordan algebra $J_{3} ^{\mathbb{O}_S }$
over the split octonions $\mathbb{O}_S$ \cite{Gunaydin:2000xr}.

%%%%%%%%%%%%%%%%%%%%%%%%%%%%%%%%%%%%%%%%%%%%%%%%%%%%%%%%%%%%%%%%%%%%%%
%%%%%%%%%%%%%%%%%%%%%%%%%%%%%%%%%%%%%%%%%%%%%%%%%%%%%%%%%%%%%%%%%%%%%%
%%%%%%%%%%%%%%%%%%%%%%%%%%%%%%%%%%%%%%%%%%%%%%%%%%%%%%%%%%%%%%%%%%%%%%
%%%%%%%%%%%%%%%%%%%%%%%%%%%%%%%%%%%%%%%%%%%%%%%%%%%%%%%%%%%%%%%%%%%%%%
%%%%%%%%%%%%%%%%%%%%%%%%%%%%%%%%%%%%%%%%%%%%%%%%%%%%%%%%%%%%%%%%%%%%%%

\section{ $E_{7(7)}$\, invariants and $\mathcal{N}=8$ black holes}
\label{sec-invariantsandbhs}

%%%%%%%%%%%%%%%%%%%%%%%%%%%%%%%%%%%%%%%%%%%%%%%%%%%%%%%%%%%%%%%%%%%%%%
%%%%%%%%%%%%%%%%%%%%%%%%%%%%%%%%%%%%%%%%%%%%%%%%%%%%%%%%%%%%%%%%%%%%%%
%%%%%%%%%%%%%%%%%%%%%%%%%%%%%%%%%%%%%%%%%%%%%%%%%%%%%%%%%%%%%%%%%%%%%%
%%%%%%%%%%%%%%%%%%%%%%%%%%%%%%%%%%%%%%%%%%%%%%%%%%%%%%%%%%%%%%%%%%%%%%
%%%%%%%%%%%%%%%%%%%%%%%%%%%%%%%%%%%%%%%%%%%%%%%%%%%%%%%%%%%%%%%%%%%%%%

\subsection{Charges}

Here we review, for the sake of completeness and as an introduction to the
constructions that we will present later, several well-known results and
concepts related to the entropy of the black holes of $\mathcal{N}=8$
supergravity.

Objects transforming in the fundamental (i.e~$\mathbf{56}$) representation,
such as the fundamental vector doublets $\mathcal{F}$ discussed before, can be
written in two different bases. In the first basis, $\mathcal{F}$ consists in
a pair of real, antisymmetric, independent tensors $F^{ij}$ and $G_{ij}\,
,\,\, i,j=1,\cdots, 8$ and the action of $E_{7(7)}$\, embedded into $Sp(56,
\mathbb{R})$ is given by
\begin{eqnarray}
\label{symplecticSmall}
\delta \mathcal{F}
=
\delta \left(
\begin{array}{cc}
F^{ij} \\
G_{ij} \\
\end{array}
\right) 
=
\left(
\begin{array}{cc}
2\Lambda ^{[i}{}_{k} \delta^{j]}{}_{l}& \Sigma^{ijkl} \\
\Sigma_{ijkl} &   2\Lambda _{[i}{}^{k} \delta_{j]}{}^{l}  \\
\end{array}
\right)  
 \left(
\begin{array}{cc}
F^{kl} \\
G_{kl} \\
\end{array}
\right)\, .
\end{eqnarray}
where the $\Lambda^{i}{}_{j}$ are infinitesimal transformations of the
maximal, (non-compact) subgroup $SL(8,\mathbb{R})$
(i.e.~$\Lambda^{i}{}_{i}=0$) and where the off-diagonal infinitesimal
parameters satisfy
\begin{equation}
\Sigma^{ijkl} = \tfrac{1}{4!}\varepsilon^{ijklmnpq}\Sigma_{mnpq}\, .  
\end{equation}
$F^{ij}$ and $G_{ij}$ transform separately contravariantly and covariantly,
respectively, in the $\mathbf{28}$ of $SL(8,\mathbb{R})$. Together, they
transform as components of a symplectic vector in the $\mathbf{56}$ of
$E_{7(7)}\subset Sp(56,\mathbb{R})$.

In the second basis $\mathcal{F}$ is written as a complex, antisymmetric
tensor with components $\mathcal{F}_{AB}$ transforming infinitesimally under
$E_{7(7)}$ as
\begin{equation}
\label{eq:E77transformSU8basis}
\begin{array}{rcl}
\delta \overline{\mathcal{F}}^{AB} 
& =  &
+2\Lambda^{[A|}{}_{C}
\overline{\mathcal{F}}^{\, C|B]} +\overline{\Sigma}^{ABCD}\mathcal{F}_{AB}\, ,
\\
& & \\
\delta \mathcal{F}_{AB} 
& =  &
-2\Lambda^{C|}{}_{[A|}
\mathcal{F}_{C|B]} +\Sigma_{ABCD}\overline{\mathcal{F}}^{\,AB}\, ,
\\
\end{array}
\end{equation}
Here, the $\Lambda^{A}{}_{B}$ are infinitesimal transformations of the
maximal, compact subgroup $SU(8)$ (i.e.~$\Lambda^{I}{}_{I}=0$) and where the
off-diagonal infinitesimal parameters $\Sigma_{ABCD}$ satisfy the complex
self-duality condition
\begin{equation}
\label{eq:complexselfdualitySigma}
\overline{(\Sigma_{ABCD})}\equiv \overline{\Sigma}^{ABCD} = 
\tfrac{1}{4!}\varepsilon^{ABCDEFGH}\Sigma_{EFGH}\, .  
\end{equation}
Thus $\mathcal{F}_{AB}$ transforms in the $\mathbf{28}$ of $SU(8)$. 

The $\Lambda$ and $\Sigma$ transformations in one basis are a combination of
the $\Lambda$ and $\Sigma$ transformations of the other one, but it is a
remarkable fact that, algebraically, they appear in a very similar way in both
cases. This means that, if we construct an $E_{7(7)}$ invariant in the
$SL(8,\mathbb{R})$ basis, we can immediately write another one (not
necessarily equivalent) in the $SU(8)$ basis by formally replacing everywhere
the components $G_{ij}$ by $\mathcal{F}_{AB}$ and the components $F^{ij}$ by
$\overline{\mathcal{F}}^{\, AB}$.

The relation between the components of $X$ in both bases is
\begin{equation}
\label{eq:basischange}
\overline{\mathcal{F}}^{\, AB} \equiv \tfrac{1}{4\sqrt{2}}\left(F^{ij}-iG_{ij} \right)
\Gamma^{ij}{}_{AB}\, , 
\end{equation}
where the $\Gamma^{ij}$s are the $SO(8)$ gamma matrices.

Let us consider the charges of the theory. Associated to the electric and
magnetic field strengths $F^{ij}$ and $G_{ij}$ we have as many magnetic and
electric charges resp.~$p^{ij}$ and $q_{ij}$ defined by
\begin{equation}
p^{ij} \equiv \int_{S^{2}_{\infty}}F^{ij}\, ,
\hspace{1cm}   
q_{ij} \equiv \int_{S^{2}_{\infty}}G_{ij}\, ,
\end{equation}
that we can combine into a real, symplectic, charge vector
\begin{equation}
\mathcal{Q} \equiv   
\left(
  \begin{array}{c}
  p^{ij} \\ q_{ij} \\  
  \end{array}
\right)\,  , 
\end{equation}
which will transform in the $\mathbf{56}$ $E_{7(7)}\subset
Sp(56,\mathbb{R}))$. This object occurs naturally in the $SL(8,\mathbb{R})$
basis, but we can rewrite it in the $SU(8)$ basis using the above formulae
\begin{equation}
\label{eq:QAB}
\overline{\mathcal{Q}}^{\, AB} \equiv \tfrac{1}{4\sqrt{2}}\left(p^{ij}-iq_{ij} \right)
\Gamma^{ij}{}_{AB}\, . 
\end{equation}
This object is sometimes written in the literature as $Z^{AB}$, which seems to
suggest that it is the central charge of the theory, which it is not since, in
particular, it is moduli-independent (hence the change of notation). The
central charge of the theory is a moduli-dependent quantity given by
\begin{equation}
\label{eq:centralchargedef}
\mathcal{Z}_{IJ}(\phi,Q)  
\equiv
\langle \mathcal{V}_{IJ}\mid \mathcal{Q} \rangle
=
\tfrac{1}{2}\left(p^{ij}h_{ij\, IJ} -q_{ij}f^{ij}{}_{IJ}\right)\, , 
\end{equation}
and only transforms under induced $SU(8)$ transformations.  Its value at
spatial infinity is the (magnetic) charge of the graviphoton
\begin{equation}
\mathcal{Z}_{IJ\, \infty}=\mathcal{Z}_{IJ}(\phi_{\infty},Q)  
= 
\int_{S^{2}_{\infty}}T_{IJ}\, .
\end{equation}

%%%%%%%%%%%%%%%%%%%%%%%%%%%%%%%%%%%%%%%%%%%%%%%%%%%%%%%%%%%%%%%%%%%%%%
%%%%%%%%%%%%%%%%%%%%%%%%%%%%%%%%%%%%%%%%%%%%%%%%%%%%%%%%%%%%%%%%%%%%%%
%%%%%%%%%%%%%%%%%%%%%%%%%%%%%%%%%%%%%%%%%%%%%%%%%%%%%%%%%%%%%%%%%%%%%%
%%%%%%%%%%%%%%%%%%%%%%%%%%%%%%%%%%%%%%%%%%%%%%%%%%%%%%%%%%%%%%%%%%%%%%
%%%%%%%%%%%%%%%%%%%%%%%%%%%%%%%%%%%%%%%%%%%%%%%%%%%%%%%%%%%%%%%%%%%%%%

\subsection{Invariants}

Let us now consider the invariants that can be constructed with the
charges. The symplectic product of two fundamentals in the real basis
\begin{equation}
\label{eq:simproduct}
\langle  \mathcal{Q}_{1} \mid  \mathcal{Q}_{2} \rangle  
= -\tfrac{1}{2}
\left[
\mathrm{Tr}_{SL(8,\mathbb{R})} \left(p_{2}\cdot q_{1}\right) 
-\mathrm{Tr}_{SL(8,\mathbb{R})} \left(p_{1}\cdot q_{2}\right) 
\right]\, ,
\end{equation}
where
\begin{equation}
\mathrm{Tr}_{SL(8,\mathbb{R})} (p\cdot q)
\equiv 
p^{ij}q_{ji}\, ,  
\end{equation}
is automatically $E_{7(7)}$ invariant since $E_{7(7)}$ acts as a subgroup of
$Sp(56,\mathbb{R})$ but it vanishes identically for a single charge
$\mathcal{Q}_{1}= \mathcal{Q}_{2}$ due to its antisymmetry.  There are no
other moduli-independent quadratic invariants.

Let us consider the quartic invariants of a single fundamental
$\mathcal{Q}$. Cartan's quartic $E_{7(7)}$\, invariant $J_{4}(\mathcal{Q})$
\cite{CARTAN} is given in the $SL(8,\mathbb{R})$ real basis by
\begin{equation}
 \label{Cartan} 
 \boxed{
J_{4}(\mathcal{Q})
= 
\mathrm{Tr}_{SL(8,\mathbb{R})} ( p\cdot q \cdot p\cdot q )
-\tfrac{1}{4} \left[\mathrm{Tr}_{SL(8,\mathbb{R})} (p\cdot q)\right]^{2} 
+\tfrac{1}{4}\, \mathrm{Pf}\, || q || 
+\tfrac{1}{4}\, \mathrm{Pf}\, || p ||\, , 
 }
\end{equation}
where 
\begin{equation}
\mathrm{Tr}_{SL(8,\mathbb{R})}  ( p\cdot q \cdot p\cdot q )
\equiv 
p^{ij}q_{jk}p^{kl}q_{li}\, ,
\end{equation}
and where $\mathrm{Pf}$ stands for the Pfaffian of an antisymmetric matrix of
even dimension, which is the square root of the determinant:
\begin{equation}
  \begin{array}{rcccl}
\mathrm{Pf}_{SL(8,\mathbb{R})}\, || q ||  & \equiv  & (\det ||q ||)^{1/2} & = &
\tfrac{1}{4!} \varepsilon^{ijklmnpq}q_{ij}q_{kl}q_{mn}q_{pq}\, , 
\\
& & & &\\
\mathrm{Pf}_{SL(8,\mathbb{R})}\, || p ||  & \equiv  & (\det ||p ||)^{1/2} & = &
\tfrac{1}{4!} \varepsilon_{ijklmnpq}p^{ij}p^{kl}p^{mn}p^{pq}\, .\\
\end{array}
\end{equation}

The Cartan invariant $J_{4}(\mathcal{Q})$ is manifestly invariant under the
global $SL(8,\mathbb{R})$ subgroup of $E_{7(7)}$.  It is enough to check the
invariance under the off-diagonal transformations generated by the selfdual
$\Sigma$ parameters
\begin{equation}
\label{eq:sigmatrans1}
\begin{array}{rclrcl}
\delta_{\Sigma}p^{ij} 
& = &
\Sigma^{ijkl}q_{kl}\, ,
&
\delta_{\Sigma}q_{ij} 
& = &
\Sigma_{ijkl}p^{kl}\, ,
\\
\end{array}
\end{equation}
to prove the invariance under the full $E_{7(7)}$. 

The Julia-Cremmer quartic invariant $\diamondsuit (\mathcal{Q})$
\cite{Cremmer:1979up} is defined in the complex basis and can be obtained from
$J_{4}$ by following the recipe given in the paragraph below
Eq.~(\ref{eq:complexselfdualitySigma}), which explains why it is so similar to
it:
\begin{equation}
\label{eq:CremmerJulia}
\boxed{
  \diamondsuit  (\mathcal{Q})
  = 
  \mathrm{Tr}_{SU(8)} (\mathcal{Q}\overline{\mathcal{Q}}\mathcal{Q}\overline{\mathcal{Q}}) 
  -\tfrac{1}{4} \left[\mathrm{Tr}_{SU(8)} (\mathcal{Q}\overline{\mathcal{Q}}) \right]^{2} 
  +\tfrac{1}{4}\mathrm{Pf}_{SU(8)}\, ||\mathcal{Q}||
  +\tfrac{1}{4}\mathrm{Pf}_{SU(8)}\, ||\overline{\mathcal{Q}}||\, ,
}
\end{equation}
where
\begin{equation}
  \begin{array}{rcl}
\mathrm{Tr}_{SU(8)} (\mathcal{Q}\overline{\mathcal{Q}}\mathcal{Q}\overline{\mathcal{Q}}) 
& \equiv & 
\mathcal{Q}_{AB}\overline{\mathcal{Q}}^{BC}\mathcal{Q}_{CD}\overline{\mathcal{Q}}^{DE}\, ,
\\
& & \\
\mathrm{Tr}_{SU(8)} (\mathcal{Q}\overline{\mathcal{Q}})
& \equiv & 
-\mathcal{Q}_{AB}\overline{\mathcal{Q}}^{AB}\, ,
\\
& & \\
\mathrm{Pf}_{SU(8)}\, ||\mathcal{Q}||
& \equiv & 
\tfrac{1}{4!} \varepsilon^{ABCDEFGH}\mathcal{Q}_{AB}\mathcal{Q}_{CD}\mathcal{Q}_{EF}\mathcal{Q}_{GH}\, ,
\\
& & \\
\mathrm{Pf}_{SU(8)}\, ||\overline{\mathcal{Q}}||
& \equiv &
\overline{\mathrm{Pf}_{SU(8)}\, ||\mathcal{Q}||}\, ,   
\end{array}
\end{equation}
and $\mathcal{Q}_{AB}$ is the complex combination of the electric and magnetic
charges defined in Eq.~(\ref{eq:QAB}). $\diamondsuit$ is manifestly invariant
under the global $SU(8)$ subgroup of $E_{7(7)}$. Again, it is enough to check
the invariance under the off-diagonal transformations generated by the complex
self-dual $\Sigma$ parameters
\begin{equation}
\label{eq:sigmatrans2}
\begin{array}{rclrcl}
\delta_{\Sigma}\mathcal{Q}_{AB} 
& = & 
\Sigma_{ABCD}\overline{\mathcal{Q}}^{\, CD}\, ,  
& 
\delta_{\Sigma}\overline{\mathcal{Q}}^{AB} 
& = & 
\overline{\Sigma}^{ABCD}\mathcal{Q}_{\, CD}\, ,  
\end{array}
\end{equation}
to prove the invariance under the full $E_{7(7)}$, but this follows from the
invariance of $J_{4}$, as observed in the paragraph below
Eq.~(\ref{eq:complexselfdualitySigma}).

It was argued in~\cite{Cremmer:1979up} that $J_{4}(\mathcal{Q})$ and
$\diamondsuit (\mathcal{Q})$ should be proportional.  The precise relation was
established in \cite{Balasubramanian:1997az}
\begin{equation}
J_{4}(\mathcal{Q})= -\diamondsuit (\mathcal{Q})\, .
\end{equation}
The detailed proof was presented in \cite{Gunaydin:2000xr}. 

Now let us consider invariants for the central charge
$\mathcal{Z}_{IJ}(\phi,Q)$.  In this case it is enough to build
manifestly-$SU(8)$-invariant quantities. At the quadratic level there is only
one\footnote{It should be stressed that the $SU(8)$ indices $A,B,C,\ldots$
  that we have just discussed are different from the $SU(8)$ indices
  $I,J,K,\ldots$ of the $E_{7(7)}/SU(8)$ coset of scalars: the former
  transform under $E_{7(7)}$ as described above and the latter always
  transform via induced $SU(8)$ transformations. This means that, for
  instance, the $SU(8)$-invariant quadratic expression
  $\overline{\mathcal{Z}}^{\, IJ}\mathcal{Z}_{IJ}$ is $E_{7(7)}$ invariant
  (but moduli-dependent) while the manifestly $SU(8)$-invariant quadratic
  expression $\overline{\mathcal{Q}}^{\, AB}\mathcal{Q}_{AB}$ is
  moduli-independent but not fully $E_{7(7)}$ invariant.}
\begin{equation}
\mathrm{Tr}_{SU(8)}\left(\overline{\mathcal{Z}}\mathcal{Z}\right)
=\overline{\mathcal{Z}}^{\, IJ}  \mathcal{Z}_{JI}\, .
\end{equation}

As shown in \cite{Andrianopoli:1996ve}, this invariant gives the black-hole
potential of the FGK formalism \cite{Ferrara:1997tw}:
\begin{equation}
\label{eq:Vbh}
V_{\rm bh}(\phi,\mathcal{Q})
=
-\mathrm{Tr}_{SU(8)}\left(\overline{\mathcal{Z}}\mathcal{Z}\right)
=
\tfrac{1}{2}\mathcal{Q}^{T}
\mathcal{M}(\mathcal{N})\mathcal{Q}\, ,
\end{equation}
where
\begin{equation}
  \begin{array}{rcl}
\mathcal{M}(\mathcal{N})
 & \equiv & 
\left(
  \begin{array}{lr}
    (\mathfrak{I}+\mathfrak{R}\mathfrak{I}^{-1}\mathfrak{R})_{\Lambda\Sigma} &
    -(\mathfrak{R}\mathfrak{I}^{-1})_{\Lambda}{}^{\Sigma} \\
    & \\
    -(\mathfrak{I}^{-1}\mathfrak{R})^{\Lambda}{}_{\Sigma} &
    (\mathfrak{I}^{-1})^{\Lambda\Sigma} \\   
  \end{array}
\right)\, ,
\\
& & \\
\mathfrak{R}_{\Lambda\Sigma} & \equiv & \Re{\rm e}\, (\mathcal{N}_{\Lambda\Sigma})\, ,
\hspace{1cm}
\mathfrak{I}_{\Lambda\Sigma} \equiv \Im{\rm m}\, (\mathcal{N}_{\Lambda\Sigma})\, ,
\hspace{1cm}
(\mathfrak{I}^{-1})^{\Lambda\Sigma}\mathfrak{I}_{\Sigma\Gamma} = 
\delta^{\Lambda}{}_{\Gamma}\, .
\end{array}
\end{equation}

At the quartic level four $E_{7(7)}$ invariants can be constructed from
$\mathcal{Z}_{IJ}$, namely
\begin{equation}
\mathrm{Tr}_{SU(8)}\left(\mathcal{Z}\overline{\mathcal{Z}}\mathcal{Z
}\overline{\mathcal{Z}}\right)
\hspace{1cm}
\left[\mathrm{Tr}_{SU(8)}\left(\overline{\mathcal{Z}}\mathcal{Z}\right)\right]^{2}\, ,
\hspace{1cm}
\mathrm{Pf}_{SU(8)}||\mathcal{Z}||\, ,  
\end{equation}
and the complex conjugate of the latter (the other two are real). In this case
we are interested in real, moduli-independent, $E_{7(7)}$ invariants and, if
any, they should be linear combinations of those four invariants. To check the
moduli-independence we must take the derivative of the linear combination
w.r.t.~the scalar fields and then use the following identity 
\begin{equation}
\label{eq:covariant}
\mathfrak{D}\mathcal{Z}_{IJ} =
\tfrac{1}{2}\mathcal{P}_{IJKL}\overline{\mathcal{Z}}^{\, KL}\, ,  
\end{equation}
which follows from the Maurer-Cartan equations (\ref{eq:Maurer-Cartan}).  The
Vielbein components satisfy the complex self-duality constraint
\begin{equation}
\overline{(\mathcal{P}_{IJKL})}=\overline{\mathcal{P}}^{\, IJKL} = 
\tfrac{1}{4!}\varepsilon^{IJKLMNOP}\mathcal{P}_{MNOP}\, .  
\end{equation}
This constraint should be compared with
Eq.~(\ref{eq:complexselfdualitySigma}).

Observe that the covariant derivative Eq.~(\ref{eq:covariant}) of
$\mathcal{Z}_{IJ}$ has the same structure as the $\Sigma$ transformation of
$\mathcal{Q}_{AB}$ Eq.~(\ref{eq:sigmatrans2}) with the complex self-dual
parameter $\Sigma_{ABCD}$ replaced by the complex-selfdual Vielbein
$\mathcal{P}_{IJKL}$. This means that, if we construct an $E_{7(7)}$ for
$\mathcal{Q}$ in the $SL(8,\mathbb{R})$, then we can construct another in the
$SU(8)$ basis replacing $q_{ij}$ by $\mathcal{Q}_{AB}$ and $p^{ij}$ by
$\overline{\mathcal{Q}}^{AB}$ and then, replacing everywhere
$\mathcal{Q}_{AB}$ by $\mathcal{Z}_{IJ}$, with the corresponding change of
indices $A,B,C,D,\ldots$ by $I,J,K,L,\ldots$, we automatically obtain a
moduli-independent $E_{7(7)}$ invariant. In particular
\begin{equation}
\label{eq:Zinvariant}
\boxed{
\diamondsuit(\mathcal{Z})
\equiv  
\mathrm{Tr}_{SU(8)}\left(\mathcal{Z}
\overline{\mathcal{Z}}\mathcal{Z}\overline{\mathcal{Z}}\right)
-\tfrac{1}{4}\left[
\mathrm{Tr}_{SU(8)}\left(\mathcal{Z}\overline{\mathcal{Z}}\right)
\right]^{2}\, 
+\tfrac{1}{4}\left(
\mathrm{Pf}_{SU(8)}||\mathcal{Z}||+\mathrm{c.c.}
\right)\, ,  
}
\end{equation}
is manifestly $E_{7(7)}$ invariant because each term is separately invariant,
and, according to the above argument, it is also automatically
moduli-independent \cite{Ferrara:2006em}.  We have used the same symbol as for
the Cremmer-Julia invariant because, this invariant is formally identical,
although the transformation rules of the $SU(8)$ indices are completely
different.

Since this invariant is quartic in charges and moduli-independent, it has been
argued \cite{Andrianopoli:1996ve} that $\diamondsuit(\mathcal{Z})$ is also
proportional to $\diamondsuit(\mathcal{Q})$. The identity can be checked in
the vanishing scalars limit. To this (zeroth) order in scalars, the components
of the symplectic section $\mathcal{V}_{IJ}$ defined in
(\ref{eq:symplecticsection}) are given by
\begin{equation}
\label{eq:zerothorderscalars}
f^{ij}{}_{IJ}= \tfrac{i}{2\sqrt{2}}\Gamma^{ij}{}_{IJ}\, ,
\hspace{1cm}  
h_{ij\, IJ}= \tfrac{1}{2\sqrt{2}}\Gamma^{ij}{}_{IJ}\, ,
\end{equation}
where $\Gamma^{ij}{}_{IJ}$ are the $SO(8)$ gamma matrices,
and, then, the components of the central charge matrix are given by
\begin{equation}
  \mathcal{Z}_{IJ} = \tfrac{1}{4\sqrt{2}} \left( p^{ij}
    -iq_{ij}\right)\Gamma^{ij}{}_{IJ}\, ,   
\end{equation}
which is nothing but $\overline{\cal Q}^{AB}$, defined in Eq.~(\ref{eq:QAB}),
written with indices $I,J$. Then, the scalar independence of the whole
expression extends the identity to arbitrary values of the scalars.

%%%%%%%%%%%%%%%%%%%%%%%%%%%%%%%%%%%%%%%%%%%%%%%%%%%%%%%%%%%%%%%%%%%%%%
%%%%%%%%%%%%%%%%%%%%%%%%%%%%%%%%%%%%%%%%%%%%%%%%%%%%%%%%%%%%%%%%%%%%%%
%%%%%%%%%%%%%%%%%%%%%%%%%%%%%%%%%%%%%%%%%%%%%%%%%%%%%%%%%%%%%%%%%%%%%%
%%%%%%%%%%%%%%%%%%%%%%%%%%%%%%%%%%%%%%%%%%%%%%%%%%%%%%%%%%%%%%%%%%%%%%

\subsubsection{``Linearization'' of the quartic invariant}

In non-associative algebras the process of ``linearization'' (sometimes called
``polarization'') is defined as follows: given a homogeneous polynomial $p(x)$
of degree $n$, the process of ``linearization'' is designed to create a
completely symmetric multilinear polynomial $p^{\prime}(x_{1},..., x_{n})$ in
$n$ variables such that the original polynomial is recovered when all the
variables $x_{i}$ have the same value $x$, that is $p^{\prime}(x,\cdots, x)=
p(x)$. For example, the full linearization of the cube\footnote{A convention
  for products of more than two elements (right $x^{3}=xxx\equiv x(xx)$ or
  left association $x^{3}=xxx \equiv (xx)x$ is understood to have been chosen,
  the particular conventions are irrelevant in our discussion.}  $x^{3}= xxx$
of degree 3 is
\begin{equation}
  \tfrac{1}{3!}  (x_{1} x_{2} x_{3} + x_{1} x_{3} x_{2}+
  x_{2} x_{1} x_{3}+ x_{2} x_{3} x_{1}+x_{3} x_{1} x_{2} +x_{3} x_{2} x_{1})\, .
\end{equation}

Here we are interested in the invariant obtained as the linearization of the
quartic invariants. A 4-linear invariant
$q(\mathcal{Q}_{1},\mathcal{Q}_{2},\mathcal{Q}_{3},\mathcal{Q}_{4})$ such that
\begin{equation}
  J_{4}(\mathcal{Q})
  =
  q(\mathcal{Q},\mathcal{Q},\mathcal{Q},\mathcal{Q})\, ,
\end{equation}
was explicitly constructed in \cite{Faulkner} in a form based on Freudenthal
triple systems. The linearization of $J_{4}(\mathcal{Q})$ is obtained by
averaging over all the permutations of the four entries:
\begin{equation}
J^{\prime}_{4}(\mathcal{Q}_{1},\mathcal{Q}_{2},\mathcal{Q}_{3},\mathcal{Q}_{4})
= 
\tfrac{1}{4!}
\sum _{\pi \in S^{4}} 
q
(\mathcal{Q}_{1\pi},\mathcal{Q}_{2\pi},\mathcal{Q}_{3\pi},\mathcal{Q}_{4\pi})\, ,
\end{equation}
and satisfies 
\begin{equation}
J_{4}(\mathcal{Q})
=
J^{\prime}_{4}(\mathcal{Q},\mathcal{Q},\mathcal{Q},\mathcal{Q})\, .
\end{equation}

An explicit form of this quartic invariant in the $SL(8,\mathbb{R})$ basis
convenient for studies of $\mathcal{N}=8$ supergravity was given in
\cite{Gunaydin:2000xr} using the symplectic product and the existence the
triple product
\begin{equation}
\mathbf{56}\times\mathbf{56}\times\mathbf{56}\longrightarrow\mathbf{56}\, ,
\end{equation}
that we can denote by $(\mathcal{Q}_{1},\mathcal{Q}_{2},\mathcal{Q}_{3})$ and
whose explicit form can be found in \cite{Gunaydin:2000xr}. The resulting
4-linear quartic invariant is given by
\begin{equation}
q(\mathcal{Q}_{1},\mathcal{Q}_{2},\mathcal{Q}_{3},\mathcal{Q}_{4})=  
\langle (\mathcal{Q}_{1},\mathcal{Q}_{2},\mathcal{Q}_{3}) \mid \mathcal{Q}_{4}
  \rangle\, ,  
\end{equation}
and it was shown (indirectly) to agree with the one in \cite{Faulkner}. It can
be modified and, in particular, symmetrized, by adding products of symplectic
products of charge vectors, such as $\langle \mathcal{Q}_{1}\mid
\mathcal{Q}_{2}\rangle \langle \mathcal{Q}_{3}\mid
\mathcal{Q}_{4}\rangle$. The fully symmetrized invariant can be identified
with the linearization
$J^{\prime}_{4}(\mathcal{Q}_{1},\mathcal{Q}_{2},\mathcal{Q}_{3},\mathcal{Q}_{4})$
and its explicit form is
\begin{equation}
\label{eq:J4prime}
\boxed{
  \begin{array}{rcl}
J^{\prime}_{4}(\mathcal{Q}_{1},\mathcal{Q}_{2},\mathcal{Q}_{3},\mathcal{Q}_{4})
& \equiv &
\tfrac{1}{6}
\mathrm{Tr}_{SL(8,\mathbb{R})}
\left\{
 p_{1} \cdot  q_{2} \cdot p_{3} \cdot q_{4} 
+p_{1} \cdot q_{3} \cdot p_{4} \cdot q_{2}
+p_{1} \cdot q_{4} \cdot p_{2} \cdot q_{3}
+(p\leftrightarrow q)
\right\}
\\
% & & \\
% & & 
% \left.  
% +p_{2} \cdot q_{3} \cdot p_{4} \cdot q_{1}
% +p_{2} \cdot q_{1} \cdot p_{3} \cdot q_{4} 
% +p_{3} \cdot q_{1} \cdot p_{4} \cdot q_{2}
% \right\}
% \\
& & \\
& &   
-\tfrac{1}{12}
\left\{
[\mathcal{Q}_{1}\mid \mathcal{Q}_{2} ] [\mathcal{Q}_{3} \mid \mathcal{Q}_{4} ]
+
[\mathcal{Q}_{1}\mid \mathcal{Q}_{3} ] [\mathcal{Q}_{2} \mid \mathcal{Q}_{4} ]
+
[\mathcal{Q}_{1}\mid \mathcal{Q}_{4} ] [\mathcal{Q}_{2} \mid \mathcal{Q}_{3} ]
\right\}
\\
& & \\
& & 
+\tfrac{1}{4}
\left[
\mathrm{Pf}_{SL(8,\mathbb{R})}||p_{1} p_{2} p_{3} p_{4}||
+
(p\leftrightarrow q)
\right]\, ,
% +\tfrac{1}{4}
% \mathrm{Pf}_{SL(8,\mathbb{R})}||q_{1} q_{2} q_{3} q_{4}||\, ,
\end{array}
}
\end{equation}
where we have defined, for convenience, the symmetric product
\begin{equation}
[\mathcal{Q}_{1}\mid \mathcal{Q}_{2} ]
\equiv  
-\tfrac{1}{2}\mathrm{Tr}_{SL(8,\mathbb{R})}(p_{1}\cdot q_{2} +
(p\leftrightarrow q) )
=\tfrac{1}{2}\left(p_{1}{}^{ij}q_{2\, ij}+p_{2}{}^{ij}q_{1\, ij}\right)\, ,  
\end{equation}
and we are using the notation
\begin{equation}
\label{Pf}
  \begin{array}{rcl}
\mathrm{Pf}||p_{1} p_{2} p_{3} p_{4}||
& \equiv & 
\tfrac{1}{4!}
\varepsilon^{ijklmnop}
q_{1\, ij}q_{2\, kl}q_{3\, mn}q_{4\, op}\, ,
\\
& & \\
\mathrm{Pf}||q_{1} q_{2} q_{3} q_{4}||
& \equiv & 
\tfrac{1}{4!}
\varepsilon_{ijklmnop}
p_{1}^{ij}p_{2}^{kl}p_{3}^{mn}p_{4}^{op}\, .
\end{array}
\end{equation}

Observe that the first six terms are the linearization of the first term
$\mathrm{Tr}_{SL(8,\mathbb{R})} \left( p \cdot q \cdot p \cdot q \right)$ in
the Cartan invariant $J_{4}(\mathcal{Q})$ Eq.~(\ref{Cartan}), the next three
terms are the linearization of the second term in $J_{4}(\mathcal{Q})$,
$\left[\mathrm{Tr}_{SL(8,\mathbb{R})} \left( p \cdot q \right)\right]^{2}$,
and the last two are the linearization of the Pfaffians. They are
automatically symmetric, and only one term is needed for each Pfaffian.

It is important here that each of the four $\mathbf{56}$'s appears linearly in
this invariant.  By construction, each term is manifestly invariant by itself
under the 63 $SL(8,\mathbb{R})$ transformations parametrized by the
$\Lambda^{i}{}_{j}$ of (\ref{symplecticSmall}). However, the remaining 70
$E_{7(7)}$\, transformations parametrized by the off-diagonal $\Sigma_{ijkl}$
mix all the terms in
$J^{\prime}_{4}(\mathcal{Q}_{1},\mathcal{Q}_{2},\mathcal{Q}_{3},\mathcal{Q}_{4})$
in a non-trivial way, which makes the contribution of all the terms necessary
to have complete $E_{7(7)}$\, symmetry.

From the point of view of the $E_{7(7)}$\, symmetry, the fact that all these
terms have to cooperate to provide an invariant can be established, of course,
by a brute force computation, using (\ref{symplecticSmall}). However, since
the structure of the invariant has a significant impact on the UV finiteness
of $\mathcal{N}=8$ supergravity, it may be useful to explain the octonionic
origin of this $E_{7(7)}$\, invariant in the context of the non-Euclidean
Jordan algebras, following \cite{Gunaydin:2000xr,Gunaydin:2009zza} and
\cite{Borsten:2008wd}. For this purpose we need to change the basis from the
Cartan one, used in (\ref{eq:J4prime}) to either the Freudenthal/Jordan or the
Fano basis. We present some useful information on this in
Appendix~\ref{ap-oct}.

It is evident that following the rule explained in the paragraph above
Eq.~(\ref{eq:Zinvariant}) we can obtain the Cremmer-Julia invariant
$\diamondsuit (\mathcal{Q})$ (\ref{eq:CremmerJulia}) from the Cartan invariant
$J_{4}(\mathcal{Q})$ (\ref{Cartan}) and using it we can get the linearization
of the Cremmer-Julia invariant
$\diamondsuit^{\prime}(\mathcal{Q}_{1},\mathcal{Q}_{2},\mathcal{Q}_{3},\mathcal{Q}_{4})$
from the linearization of the Cartan invariant given in
Eq.~(\ref{eq:J4prime}):
\begin{equation}
\label{eq:diamondprime}
\boxed{
  \begin{array}{rcl}
\diamondsuit^{\prime}(\mathcal{Q}_{1},\mathcal{Q}_{2},\mathcal{Q}_{3},\mathcal{Q}_{4})
& \equiv &
\tfrac{1}{6}
\mathrm{Tr}_{SU(8)}
\left\{
 \overline{\cal Q}_{1} \cdot  {\cal Q}_{2} \cdot \overline{\cal Q}_{3} \cdot {\cal Q}_{4} 
+\overline{\cal Q}_{1} \cdot {\cal Q}_{3} \cdot \overline{\cal Q}_{4} \cdot {\cal Q}_{2}
+\overline{\cal Q}_{1} \cdot {\cal Q}_{4} \cdot \overline{\cal Q}_{2} \cdot {\cal Q}_{3}
+\mathrm{c.c.}
\right\}
% \right.
% \\
% & & \\
% & & 
% \left.  
% +\overline{\cal Q}_{2} \cdot {\cal Q}_{3} \cdot \overline{\cal Q}_{4} \cdot {\cal Q}_{1}
% +\overline{\cal Q}_{2} \cdot {\cal Q}_{1} \cdot \overline{\cal Q}_{3} \cdot {\cal Q}_{4} 
% +\overline{\cal Q}_{3} \cdot {\cal Q}_{1} \cdot \overline{\cal Q}_{4} \cdot {\cal Q}_{2}
% \right\}
\\
& & \\
& &   
-\tfrac{1}{12}
\left\{
[\mathcal{Q}_{1}\mid \mathcal{Q}_{2} ] [\mathcal{Q}_{3} \mid \mathcal{Q}_{4} ]
+
[\mathcal{Q}_{1}\mid \mathcal{Q}_{3} ] [\mathcal{Q}_{2} \mid \mathcal{Q}_{4} ]
+
[\mathcal{Q}_{1}\mid \mathcal{Q}_{4} ] [\mathcal{Q}_{2} \mid \mathcal{Q}_{3} ]
\right\}
\\
& & \\
& & 
+\tfrac{1}{4}
\left[
\mathrm{Pf}_{SU(8)}
||\overline{\cal Q}_{1} \overline{\cal Q}_{2} \overline{\cal Q}_{3} \overline{\cal Q}_{4}||
+
\mathrm{c.c.}
\right]\, ,
% +\tfrac{1}{4}
% \mathrm{Pf}_{SU(8)}
% ||{\cal Q}_{1} {\cal Q}_{2} {\cal Q}_{3} {\cal Q}_{4}||\, ,
\end{array}
}
\end{equation}
where the Pfaffians and traces have the obvious expressions and the symmetric
products are given by
\begin{equation}
[\mathcal{Q}_{1}\mid \mathcal{Q}_{2} ]  
=
-\tfrac{1}{2}\mathrm{Tr}_{SU(8)}\left[\overline{\cal Q}_{1}\cdot
  \mathcal{Q}_{2}\right] +\mathrm{c.c.}
=
\tfrac{1}{2}\left[\overline{\cal Q}_{1}{}^{AB}\mathcal{Q}_{2\, AB}
+\mathrm{c.c.}\right]\, .
\end{equation}

Finally, we can also linearize the moduli-independent $E_{7(7)}$ invariant of
the central charge (\ref{eq:Zinvariant}), but the linearization procedure does
not necessarily guarantee moduli-independence. However, according to the
observation in the paragraph above (\ref{eq:Zinvariant}), if we replace in the
linearization of the Cremmer-Julia invariant (\ref{eq:diamondprime})
$\mathcal{Q}_{i\, AB}$ by $\mathcal{Z}_{i\, IJ}$, we immediately get a
moduli-independent $E_{7(7)}$ invariant expression. Nevertheless, there is an
important subtlety we must pay attention to: the contraction of ``local''
$SU(8)$ indices $I,J,\ldots$ is $SU(8)$ and $E_{7(7)}$ invariant if and only
if the scalars that occur in the $SU(8)$ matrices are functions of the same
variables. For example
\begin{equation}
\label{eq:noinvariance}
\mathcal{Z}_{1\, IJ}\overline{\cal Z}_{2}{}^{IJ}
=
\mathcal{Z}_{IJ}[\phi(x_{1}),\mathcal{Q}_{1}]\, \overline{\cal
  Z}^{IJ}[\phi(x_{2}),\mathcal{Q}_{2}]\, ,  
\end{equation}
will only be invariant if $\phi(x_{1})=\phi(x_{2})$, which, generically,
requires that $x_{1}=x_{2}$.

When we discuss the construction of $E_{7(7)}$ invariants for amplitudes from
graviphoton field strengths, this observation will play an important role.

%%%%%%%%%%%%%%%%%%%%%%%%%%%%%%%%%%%%%%%%%%%%%%%%%%%%%%%%%%%%%%%%%%%%%%
%%%%%%%%%%%%%%%%%%%%%%%%%%%%%%%%%%%%%%%%%%%%%%%%%%%%%%%%%%%%%%%%%%%%%%
%%%%%%%%%%%%%%%%%%%%%%%%%%%%%%%%%%%%%%%%%%%%%%%%%%%%%%%%%%%%%%%%%%%%%%
%%%%%%%%%%%%%%%%%%%%%%%%%%%%%%%%%%%%%%%%%%%%%%%%%%%%%%%%%%%%%%%%%%%%%%
%%%%%%%%%%%%%%%%%%%%%%%%%%%%%%%%%%%%%%%%%%%%%%%%%%%%%%%%%%%%%%%%%%%%%%

\subsection{Black-hole areas}

According to the general results of \cite{Ferrara:1997tw}, the area of the
extremal $\mathcal{N}=8$ black holes is given by the value of the black-hole
potential Eq.~(\ref{eq:Vbh}) on the horizon, where the scalars take values
that only depend on the charges\footnote{At least, for the supersymmetric
  ones.}. The resulting expression $V_{\rm bh}(\phi_{\rm h},\mathcal{Q})$ only
depends on the charges and is, by construction, invariant under $E_{7(7)}$
duality transformations.

It was argued in \cite{Kallosh:1996uy} that this area is actually proportional
to the square root of the Cartan-Julia-Cremmer quartic invariant
\begin{equation} 
\label{usymmetry}
A =  4\pi \sqrt{|\diamondsuit (\mathcal{Q}) |}\, ,
\end{equation}
which requires the highly non-trivial identity
\begin{equation}
V_{\rm bh}(\phi_{\rm  h},\mathcal{Q})
=
-\mathrm{Tr}_{SU(8)}\left(\overline{\mathcal{Z}}_{\rm h}\mathcal{Z}_{\rm h}\right)
=
\tfrac{1}{2}\mathcal{Q}^{T}
\mathcal{M}(\mathcal{N})_{\rm h}\mathcal{Q}
=
4\pi \sqrt{|\diamondsuit (\mathcal{Q}) |}\, .
\end{equation}
%

%%%%%%%%%%%%%%%%%%%%%%%%%%%%%%%%%%%%%%%%%%%%%%%%%%%%%%%%%%%%%%%%%%%%%%
%%%%%%%%%%%%%%%%%%%%%%%%%%%%%%%%%%%%%%%%%%%%%%%%%%%%%%%%%%%%%%%%%%%%%%
%%%%%%%%%%%%%%%%%%%%%%%%%%%%%%%%%%%%%%%%%%%%%%%%%%%%%%%%%%%%%%%%%%%%%%
%%%%%%%%%%%%%%%%%%%%%%%%%%%%%%%%%%%%%%%%%%%%%%%%%%%%%%%%%%%%%%%%%%%%%%
%%%%%%%%%%%%%%%%%%%%%%%%%%%%%%%%%%%%%%%%%%%%%%%%%%%%%%%%%%%%%%%%%%%%%%
%%%%%%%%%%%%%%%%%%%%%%%%%%%%%%%%%%%%%%%%%%%%%%%%%%%%%%%%%%%%%%%%%%%%%%

\subsection{$E_{7(7)}$ invariants in 2-center $\mathcal{N}=8$ solutions }

In the context of 2-center extremal black-hole solutions there was a complete
analysis in \cite{Andrianopoli:2011gy} of the possible $E_{7(7)}$ invariants
depending on the two fundamentals $\mathcal{Q}_{1}$ and $\mathcal{Q}_{2}$ that
characterize each black hole. the analysis takes into account the
\textit{horizontal symmetry group} $SL (2, \mathbb{R})$, which rotates the two
fundamentals. Seven of them, which are irreducible, we describe below.
These invariants help to study different physical properties, such as marginal
stability and split attractor flow solutions with two dyonic black-hole charge
vectors.

There are two invariants which are antisymmetric in $ \mathcal{Q}_{1}$ and $
\mathcal{Q}_{2}$.  One is the quadratic symplectic invariant $\langle
\mathcal{Q}_{1} \mid \mathcal{Q}_{2} \rangle$ and the other one is sextic.
The quadratic one is related to the total angular momentum of the 2-center
solution: in $\mathcal{N}=2$ theories the angular momentum of 2-center
solutions is given by the symplectic product of the charge-vectors
$\mathcal{Q}_{1,2}$ of the two centers (see, \textit{e.g.},
\cite{Bellorin:2006xr})
\begin{equation}
  \vec{J}  = \langle  \mathcal{Q}_{1} \mid  \mathcal{Q}_{2} \rangle 
  \frac{(\vec{x}_{1}-\vec{x}_{2})}{|\vec{x}_{1}-\vec{x}_{2}|}\, ,
\end{equation}
where $\vec{x}_{1,2}$ are the ``positions'' of the horizons of the two black
holes in the transverse Euclidean 3-dimensional space.  The
Dirac-Schwinger-Zwanziger quantization condition can be applied immediately to
it, implying the quantization of the total angular momentum. The same
expression must be valid in $\mathcal{N}=8$ supergravity\footnote{It follows
  from Eq.~(4.40) of \cite{Meessen:2010fh}, which is identical to the equation
  that appears in $\mathcal{N}=2$.}.  
  
The are also five different $E_{7(7)}$ invariants depending on $
\mathcal{Q}_{1}$ and $ \mathcal{Q}_{2}$ so that at $\mathcal{Q}_{1}=
\mathcal{Q}_{2}$ they all reduce to the familiar quartic invariant. They are
usually described in terms of the so-called $\mathbb{K}$-tensor which is
totally symmetric in its 4 symplectic indices\footnote{each symplectic index
  $M,N,\ldots$ is equivalent to a pair of indices $\Lambda,\Sigma,\ldots $
  each of which is equivalent, in turn, to an antisymmetric pair of
  $SL(8,\mathbb{R})$ indices $[ij]$.}
\begin{equation}
\mathbb{K}_{MNPQ}= \mathbb{K}_{(MNPQ)}\, ,
\end{equation}
and can be defined by its contraction with four different fundamentals:
\begin{equation}
\label{us}
\boxed{
 \mathbb{K}_{MNPQ} \mathcal{Q}_{1}{}^{M} \mathcal{Q}_{2}{}^{N}
 \mathcal{Q}_{3}{}^{P} \mathcal{Q}_{4}{}^{Q}
\equiv
J^{\prime}_{4}(\mathcal{Q}_{1},\mathcal{Q}_{2},\mathcal{Q}_{3},\mathcal{Q}_{4})\,
.
}
\end{equation}

The five quartic invariants relevant for the 2-center solution listed in
\cite{Andrianopoli:2011gy} are, then
\begin{equation}
  \begin{array}{rcl}
I_{+2}
& = & 
 \mathbb{K}_{MNPQ} \mathcal{Q}_{1}{}^{M} \mathcal{Q}_{1}{}^{N}
 \mathcal{Q}_{1}{}^{P} \mathcal{Q}_{1}{}^{Q}
=
J^{\prime}_{4}(\mathcal{Q}_{1},\mathcal{Q}_{1},\mathcal{Q}_{1},\mathcal{Q}_{1})
=
J_{4}(\mathcal{Q}_{1})\, ,
\\
& & \\
I_{+1}
& = & 
 \mathbb{K}_{MNPQ} \mathcal{Q}_{1}{}^{M} \mathcal{Q}_{1}{}^{N}
 \mathcal{Q}_{1}{}^{P} \mathcal{Q}_{2}{}^{Q}
=
J^{\prime}_{4}(\mathcal{Q}_{1},\mathcal{Q}_{1},\mathcal{Q}_{1},\mathcal{Q}_{2})\, ,
\\
& & \\
I_{0}
& = & 
 \mathbb{K}_{MNPQ} \mathcal{Q}_{1}{}^{M} \mathcal{Q}_{1}{}^{N}
 \mathcal{Q}_{2}{}^{P} \mathcal{Q}_{2}{}^{Q}
=
J^{\prime}_{4}(\mathcal{Q}_{1},\mathcal{Q}_{1},\mathcal{Q}_{2},\mathcal{Q}_{2})\, ,
\\
& & \\
I_{-1}
& = & 
\mathbb{K}_{MNPQ} \mathcal{Q}_{1}{}^{M} \mathcal{Q}_{2}{}^{N}
 \mathcal{Q}_{2}{}^{P} \mathcal{Q}_{2}{}^{Q}
=
J^{\prime}_{4}(\mathcal{Q}_{1},\mathcal{Q}_{2},\mathcal{Q}_{2},\mathcal{Q}_{2})\, ,
\\
& & \\
I_{-2}
& = & 
\mathbb{K}_{MNPQ} \mathcal{Q}_{2}{}^{M} \mathcal{Q}_{2}{}^{N}
\mathcal{Q}_{2}{}^{P} \mathcal{Q}_{2}{}^{Q}
=
J^{\prime}_{4}(\mathcal{Q}_{2},\mathcal{Q}_{2},\mathcal{Q}_{2},\mathcal{Q}_{2})\, ,
=
J_{4}(\mathcal{Q}_{2})\, .
\end{array}
\end{equation}

The explicit expression in (\ref{eq:J4prime}) detailing (\ref{us}) will prove
very useful, since it will allow us to deduce the specific properties of the
four-linear invariant and to compare with amplitudes. These properties can be
traced to the fact that in groups of type $E7$ the quartic invariant is not a
perfect square, whereas in the degenerate groups of type $E7$ they form a
perfect square, see \cite{Brown,Ferrara:2011dz} for more details.

%%%%%%%%%%%%%%%%%%%%%%%%%%%%%%%%%%%%%%%%%%%%%%%%%%%%%%%%%%%%%%%%%%%%%%
%%%%%%%%%%%%%%%%%%%%%%%%%%%%%%%%%%%%%%%%%%%%%%%%%%%%%%%%%%%%%%%%%%%%%%
%%%%%%%%%%%%%%%%%%%%%%%%%%%%%%%%%%%%%%%%%%%%%%%%%%%%%%%%%%%%%%%%%%%%%%
%%%%%%%%%%%%%%%%%%%%%%%%%%%%%%%%%%%%%%%%%%%%%%%%%%%%%%%%%%%%%%%%%%%%%%
%%%%%%%%%%%%%%%%%%%%%%%%%%%%%%%%%%%%%%%%%%%%%%%%%%%%%%%%%%%%%%%%%%%%%%
%%%%%%%%%%%%%%%%%%%%%%%%%%%%%%%%%%%%%%%%%%%%%%%%%%%%%%%%%%%%%%%%%%%%%%

\section{New four-linear $E_{7(7)}$\, invariants  and 4-vector amplitudes}
\label{sec-invariantsandamplitudes}

For the $E_{7(7)}$-symmetric quartic in vectors local action we need 4
different fundamental $\mathbf{56}$ depending on 4 different
momenta\footnote{As discussed above Eq.~(\ref{eq:noinvariance}), in princple
  we cannot use graviphoton field strengths depending on different momenta to
  construct $E_{7(7)}$ invariants. However, in the vanishing scalar limit the
  expressions that we will obtain can be reinterpreted in terms of graviphoton
  field strengths, see Eq.~(\ref{eq:TvsF}).}
\begin{equation}
\label{4of 56}
\overline{\mathcal{F}}^{\, AB}{}_{\mu\nu}(p_{I}) \equiv 
\tfrac{1}{4\sqrt{2}}\left(F^{ij}(p_{I})-iG_{ij}(p_{I}) \right)
\Gamma^{ij}{}_{AB}\, ,
\hspace{0.5cm}
I=1,\cdots , 4\, , 
\hspace{0.5cm}
p_{1}+p_{2}+p_{3}+p_{4}=0\, .
\end{equation}
where $i, j=1,\cdots , 8\, , \hspace{0.5cm} A,B= 1\, ,\cdots , 8$.

Using the linearization of the Cartan invariant (\ref{eq:J4prime}) we can
construct an $E_{7(7)}$-invariant Lorentz tensor with 4 pairs of antisymmetric
indices depending on the 4 momenta $p_{I}$:
\begin{equation}
\label{J}
\diamondsuit_{(8)}\equiv
\diamondsuit_{(8)\, [\mu_{1}\nu_{1}][\mu_{2}\nu_{2}][\mu_{3}\nu_{3}][\mu_{4}\nu_{4}]} 
(p_{1}, p_{2}, p_{3}, p_{4}) \equiv \diamondsuit^{\prime}\,
[\mathcal{F}_{\mu_{1}\nu_{1}}(p_{1}), \mathcal{F}_{\mu_{2}\nu_{2}}(p_{2}),
\mathcal{F}_{\mu_{3}\nu_{3}}(p_{3}), \mathcal{F}_{\mu_{4}\nu_{4}}(p_{4}) ]\, .
\end{equation}
It remains to make this expression Lorentz invariant.

%%%%%%%%%%%%%%%%%%%%%%%%%%%%%%%%%%%%%%%%%%%%%%%%%%%%%%%%%%%%%%%%%%%%%%
%%%%%%%%%%%%%%%%%%%%%%%%%%%%%%%%%%%%%%%%%%%%%%%%%%%%%%%%%%%%%%%%%%%%%%
%%%%%%%%%%%%%%%%%%%%%%%%%%%%%%%%%%%%%%%%%%%%%%%%%%%%%%%%%%%%%%%%%%%%%%
%%%%%%%%%%%%%%%%%%%%%%%%%%%%%%%%%%%%%%%%%%%%%%%%%%%%%%%%%%%%%%%%%%%%%%
%%%%%%%%%%%%%%%%%%%%%%%%%%%%%%%%%%%%%%%%%%%%%%%%%%%%%%%%%%%%%%%%%%%%%%

\subsection{Taking care of Lorentz invariance}

To make (\ref{J}) Lorentz invariant we must contract the 4 pairs of
antisymmetric Lorentz indices with another scalar- and vector-independent Lorentz
tensor. Which tensor should we use?

We may view the new duality-invariant Born-Infeld models with higher
derivatives \cite{Chemissany:2011yv} as the reduction from $E_{7(7)}$\,
symmetry to $U(1)$ and, in the corresponding manifestly $U(1)$-invariant
initial source of deformation, the 4 pairs of antisymmetric Lorentz indices
are contracted with the $t^{(8)}$ tensor. This tensor, whose definition and
properties are described in Appendix~\ref{app-t8}, is totally symmetric in
four pairs of antisymmetric indices. This fact suggests that we must use
$t^{(8)}$.
In such case, the expression which is both Lorentz- as well as $E_{7(7)}$-
invariant is
\begin{equation}
\label{O}
{\cal I}_{f} \equiv t^{(8) }\cdot \diamondsuit^{\prime}_{(8)}\;  f(s, t, u) 
\end{equation}
where $f(s, t, u)$ is an appropriate function of the Mandelstam variables and 
\begin{equation}
\label{inv}
t^{(8) }\cdot \diamondsuit_{(8)}
\equiv 
t^{(8) abcd}
\diamondsuit_{(8)\, abcd}\, ,
\end{equation}
where we have replaced each pair of Lorentz indices by a single Latin index
$a,b,\ldots$, a notation explained in Appendix~\ref{app-t8}.  It is of crucial
importance here that the function $f(s,t,u)$ only occurs as a common factor of
all the terms in $\diamondsuit_{(8)}$, because, otherwise, the $E_{7(7)}$\,
symmetry would be broken. Thus, (\ref{O}) is our candidate for to manifestly
$E_{7(7)}$-invariant source of deformation.

The detailed form of the invariant (\ref{O}) is complicated. The expression
can be simplified using the following property, derived in
Appendix~\ref{app-t8}:
\begin{eqnarray}
t^{(8)}{}_{abcd}A^{a}B^{b}C^{c}D^{d}
& = &   
4\textrm{Tr}_{L}(A^{+}B^{+})\textrm{Tr}_{L}(C^{-}D^{-})
+4\textrm{Tr}_{L}(A^{+}C^{+})\textrm{Tr}_{L}(B^{-}D^{-})
\nonumber \\
& & \nonumber \\
& & 
+4\textrm{Tr}_{L}(A^{+}D^{+})\textrm{Tr}(B^{-}C^{-})
+(+\leftrightarrow -)\, .
\end{eqnarray}
where $\mathrm{Tr}_{L}$ stands for the trace over the Lorentz indices. Observe
that only terms with the contraction of two selfdual nd two antiselfdual
tensors occur.

%%%%%%%%%%%%%%%%%%%%%%%%%%%%%%%%%%%%%%%%%%%%%%%%%%%%%%%%%%%%%%%%%%%%%%
%%%%%%%%%%%%%%%%%%%%%%%%%%%%%%%%%%%%%%%%%%%%%%%%%%%%%%%%%%%%%%%%%%%%%%
%%%%%%%%%%%%%%%%%%%%%%%%%%%%%%%%%%%%%%%%%%%%%%%%%%%%%%%%%%%%%%%%%%%%%%
%%%%%%%%%%%%%%%%%%%%%%%%%%%%%%%%%%%%%%%%%%%%%%%%%%%%%%%%%%%%%%%%%%%%%%
%%%%%%%%%%%%%%%%%%%%%%%%%%%%%%%%%%%%%%%%%%%%%%%%%%%%%%%%%%%%%%%%%%%%%%

\subsection{Comparison of the unconstrained $E_{7(7)}$\, 
invariants with the counterterms and 4-vector amplitudes }

The structure of the UV-divergent 4-vector amplitude is obtained by using the
linear twisted self-duality constraint on the manifestly $E_{7(7)}$\,
invariant source of deformation. In our case it means that we have to use a
condition $G_{\Lambda}{}^{+}=\overline{\mathcal{N}}_{\Lambda\Sigma}F^{\Sigma\,
  +}$ presented in Eq.~(\ref{eq:constraint}) and where the 28 $\Lambda
\rightarrow \, {[ij]}$.

If we are interested only in the 4-vector amplitude, we can linearize the
vector kinetic matrix $\mathcal{N}_{\Lambda\Sigma}$ 
\begin{equation}
\mathcal{N}_{\Lambda\Sigma} = -i\delta_{\Lambda\Sigma} + \mathcal{O}(\phi)\, ,  
\end{equation}
and, to the lowest order in scalars, 
\begin{equation}
G_{ij}{}^{+} \sim i F^{ij\, +}\, ,  
\end{equation}
so, in the complex $SU(8)$ basis (\ref{eq:basischange})
\begin{equation}
\overline{\cal F}^{AB} 
\sim 
\tfrac{1}{2\sqrt{2}} F^{ij\, +} \Gamma^{ij}{}_{AB}  
\sim 
-\tfrac{i}{2\sqrt{2}} G_{ij}{}^{+} \Gamma^{ij}{}_{AB}
= 
\overline{\cal  F}^{AB\, +}\, .
\end{equation}
This implies that 
\begin{equation}
\label{eq:linearforF}
\overline{\cal  F}^{AB\, -}=\mathcal{F}_{AB}{}^{+}=0\, ,
\end{equation}
which, in turn, implies that the $E_{7(7)}$ transformations
(\ref{eq:E77transformSU8basis}) reduce to global $SU(8)$ transformations
\begin{equation}
\label{eq:E77transformSU8basisreduced}
\begin{array}{rcl}
\delta \overline{\mathcal{F}}^{AB\, +} 
& =  &
+2\Lambda^{[A|}{}_{C}
\overline{\mathcal{F}}^{\, C|B]\, +}\, ,
\\
& & \\
\delta \mathcal{F}_{AB}{}^{-} 
& =  &
-2\Lambda^{C|}{}_{[A|}
\mathcal{F}_{C|B]}{}^{-}, .
\\
\end{array}
\end{equation}

Furthermore, using the expression of the the components of the symplectic
section $\mathcal{V}_{IJ}$ in this limit, given in
Eq.~(\ref{eq:zerothorderscalars}), we find 
\begin{equation}
T_{IJ} \sim   \tfrac{1}{4\sqrt{2}} \left( F^{ij} -iG_{ij}\right)\Gamma^{ij}{}_{IJ}\, ,
\end{equation}
which can be compared with (\ref{eq:basischange}). Then, at this order, taking
into account the above results, the ``local'', lower $IJ$ and ``global'' upper
$AB$ indices can be identified and, in particular,
\begin{equation}
\label{eq:TvsF}
T_{IJ} \sim   \tfrac{1}{2\sqrt{2}} F^{ij\, +} \Gamma^{ij}{}_{AB}  \sim
\overline{\mathcal{F}}^{AB\, +}\, ,\,\,\,\,
\Rightarrow\,\,\,\,
T_{IJ}{}^{+} \rightarrow \overline{\cal F}^{AB\, +}\, .
\end{equation}

The consistency of this identification is guaranteed by the comparison between
the constraints (\ref{linear}) and (\ref{eq:linearforF}) and by the fact that,
at this order in scalars, $\overline{\cal F}^{AB\, +}$ only transforms under
global $SU(8)$ transformations (\ref{eq:E77transformSU8basisreduced}), just
like the graviphoton (to this order in scalars). Observe that this implies
that, again to this order, expressions like
\begin{equation}
  \begin{array}{rcl}
\mathrm{Tr}_{SU(8)}
\left(
\mathcal{F}_{1}^{-}\overline{\mathcal{F}}_{2}^{+}
\right) 
& = & 
\mathcal{F}_{1\, AB}{}^{-}
\overline{\mathcal{F}}_{2}{}^{\, AB\, +}
\, ,  
\\
& & \\
\mathrm{Tr}_{SU(8)}
\left(
\mathcal{F}_{1}^{-}\overline{\mathcal{F}}_{2}^{+}
\mathcal{F}_{3}^{-}\overline{\mathcal{F}}_{4}^{+}
\right) 
& = & 
\mathcal{F}_{1\, AB}{}^{-}
\overline{\mathcal{F}}_{2}{}^{\, BC\, +}
\mathcal{F}_{3\, CD}{}^{-}
\overline{\mathcal{F}}_{4}{}^{\, DA\, +}\, ,
\end{array}
\end{equation}
etc.~are fully $E_{7(7)}$-invariant.

The 4-vector amplitude in $\mathcal{N}=8$ supergravity, being the MHV
amplitude, is proportional to the tree amplitude, up to a polynomial in
Mandelstam variables. Actually, according to Eq.~(3.5) of
\cite{Kallosh:2008ru}, the tree-level 4-vector amplitude is given by 
\begin{equation}
\label{eqn:4vectors}
\langle\, b_{AB}^{-}b_{CD}^{-}b_{+}^{EF}b_{+}^{GH}\, \rangle
=
-\langle12\rangle^{2}
[34]^{2}
\left[
\frac{1}{t}\delta_{AB}{}^{EF}\delta_{CD}{}^{GH}
+\frac{1}{u}\delta_{AB}{}^{GH}\delta_{CD}{}^{EF}
+\frac{1}{s}\delta_{AB}{}^{FG} \delta_{CD}{}^{HE}
\right]\, .
\end{equation}
Here $\delta_{AB}{}^{EF}$ is the antisymmetrized product of gammas with weight
one,  taking the values $0,1$ and $-1$. The first two terms correspond to
\begin{equation}
\mathrm{Tr}_{SU(8)}(\mathcal{F}^{-}_{1}\overline{\mathcal{F}}^{+}_{3})\, 
\mathrm{Tr}_{SU(8)}(\mathcal{F}^{-}_{2} \overline{\mathcal{F}}^{+}_{4})
=
\mathcal{F}_{1}^{-}{}_{AB}\overline{\mathcal{F}}_{3}^{+\, AB}
\mathcal{F}_{2}^{-}{}_{CD}\overline{\mathcal{F}}_{4}^{+\, CD}
\, ,
\end{equation}
and
\begin{equation}
\mathrm{Tr}_{SU(8)}(\mathcal{F}^{-}_{1}\overline{\mathcal{F}}^{+}_{4})\, 
\mathrm{Tr}_{SU(8)}(\mathcal{F}^{-}_{2} \overline{\mathcal{F}}^{+}_{3})
=
\mathcal{F}_{1}^{-}{}_{AB}\overline{\mathcal{F}}_{4}^{+\, AB}
\mathcal{F}_{2}^{-}{}_{CD}\overline{\mathcal{F}}_{3}^{+\, CD}
\, ,
\end{equation}
while the third term corresponds to 
\begin{equation}
\mathrm{Tr}_{SU(8)}
\left(
\mathcal{F}^{-}_{1}\overline{\mathcal{F}}^{+}_{3}
\mathcal{F}^{-}_{2}\overline{\mathcal{F}}^{+}_{4}
\right)\, .
\end{equation}

The 3-loop 4-vector part of the counterterm is related to the three-loop on
shell amplitude by multiplication by the global factor $stu$. This is shown in
Eq.~(5.1) of \cite{Kallosh:2008ru} as
\begin{equation}
\label{4vectors}
\begin{array}{rcl}
M_{\rm UV4vec}^{3}( b_{AB}^{-}, b_{CD}^{-}, b_{+}^{EF}, b_{+}^{GH})_{\rm UV} 
 & = & 
-\kappa^{4} \langle12\rangle^{2}[34]^{2}
\left[
  s\, u\, \delta_{AB}{}^{EF}\delta_{CD}{}^{GH}
+
  s\, t \,\delta_{AB}{}^{GH}\delta_{CD}{}^{EF}
  \right.
\\
& & \\
& & 
\left.
  +
  t\, u\,\delta_{AB}{}^{FG} \delta_{CD}{}^{HE}
\right]\, .
\end{array}
\end{equation}

In coordinate space the 4-vector amplitude may be given in the form
\begin{equation}
\begin{array}{rcl}
& & 
2\mathrm{Tr}_{SU(8)} 
\left(
\partial_{\mu}\partial_{\nu}\mathcal{F}^{-}{}_{\alpha\beta}
\overline{\mathcal{F}}^{+}{}_{\dot{\alpha}\dot{\beta}}
\right)
\,
\mathrm{Tr}_{SU(8)} 
\left(
\partial^{\mu}\mathcal{F}^{-\, \alpha\beta}
\partial^{\nu}\overline{\mathcal{F}}^{+\, \dot{\alpha}\dot{\beta}}
\right)
\\
& & \\
& & 
+
\mathrm{Tr}_{SU(8)} 
\left(
\partial_{\mu}\partial_{\nu}\mathcal{F}^{-}{}_{\alpha\beta}
\partial^{\mu}\overline{\mathcal{F}}^{+}{}_{\dot{\alpha}\dot{\beta}}
\mathcal{F}^{-\, \alpha\beta}
\partial^{\nu}\overline{\mathcal{F}}^{+\, \dot{\alpha}\dot{\beta}}
\right)
+\mathrm{c.c.}
\\
& & \\
& = & 
2\partial_{\mu}\partial_{\nu}\mathcal{F}^{-}{}_{AB\, \alpha\beta}
\overline{\mathcal{F}}^{+\, AB}{}_{\dot{\alpha}\dot{\beta}}
\partial^{\mu}\mathcal{F}^{-}{}_{CD}{}^{\alpha\beta}
\partial^{\nu}\overline{\mathcal{F}}^{+\, CD\, \dot{\alpha}\dot{\beta}}
\\
& & \\
& & 
+
\partial_{\mu}\partial_{\nu}\mathcal{F}^{-}{}_{AB\, \alpha\beta}
\partial^{\mu}\overline{\mathcal{F}}^{+\, BC}{}_{\dot{\alpha}\dot{\beta}}
\mathcal{F}^{-}{}_{CD}{}^{\alpha\beta}
\partial^{\nu}\overline{\mathcal{F}}^{+\, DA\, \dot{\alpha}\dot{\beta}}
+\mathrm{c.c.}
\\
& & \\
& = & 
2\mathrm{Tr}_{L}
(
\partial_{\mu}\partial_{\nu}\mathcal{F}^{-}{}_{AB}
\partial^{\mu}\mathcal{F}^{-}{}_{CD}
)
\mathrm{Tr}_{L}
\left(
\overline{\mathcal{F}}^{+\, AB}
\partial^{\nu}\overline{\mathcal{F}}^{+\, CD}
\right)
\\
& & \\
& & 
+
\mathrm{Tr}_{L}
\left(
\partial_{\mu}\partial_{\nu}\mathcal{F}^{-}{}_{AB}
\mathcal{F}^{-}{}_{CD}
\right)
\mathrm{Tr}_{L}
\left(
\partial^{\mu}\overline{\mathcal{F}}^{+\, BC}
\partial^{\nu}\overline{\mathcal{F}}^{+\, DA}
\right)
+\mathrm{c.c.}\, .
\end{array}
\end{equation}
where $\mathrm{Tr}_{L}$ stands for the trace in the Lorentz indices that are
not shown, which is close to the one given in \cite{Freedman:2011uc}.

It is clear from (\ref{4vectors}) that the terms with different $SU(8)$ index
structure have different dependence on the Mandelstam variables. This is not
possible if one uses our candidate for $E_{7(7)}$-invariant source of
deformation.  As shown in (\ref{inv}) and (\ref{O}) the 4-linear $E_{7(7)}$\,
invariant only admits a function of Mandelstam variables that appears as a
factor, common to all the $E_{7(7)}$\, and $SU(8)$ structures. Therefore the
manifestly $E_{7(7)}$\, invariant candidate for the source of deformation is
in disagreement with the counterterm\footnote{If one tries to get the
  counterterm in position space via the invariants
\begin{equation}
t^{(8)}\cdot \diamondsuit_{(8)}
\left(
\partial_{\mu}\mathcal{F},\partial_{\nu}\mathcal{F},
\partial^{\mu}\mathcal{F},\partial^{\nu}\mathcal{F}
 \right)\, ,
\hspace{.2cm}
t^{(8)}\cdot \diamondsuit_{(8)}
\left(
\partial_{\mu}\partial_{\nu}\mathcal{F},\partial^{\nu}\mathcal{F},
\partial^{\mu}\mathcal{F},\mathcal{F}
 \right)\, ,
\hspace{.2cm}
t^{(8)}\cdot \diamondsuit_{(8)}
\left(
\partial_{\mu}\partial_{\nu}\mathcal{F},\partial^{\mu}\partial^{\nu}\mathcal{F},
\mathcal{F},\mathcal{F}
 \right)\, ,
\end{equation}
one immediately sees that one finds additional terms required by invariance
under the off-diagonal part of $E_{7(7)}$ that are not present in the
counterterm.
}. In other words: the counterterm is $E_{7(7)}$ invariant, as discussed
above, but it cannot be obtained from the $E_{7(7)}$-invariant initial source
of deformation (\ref{inv}).  This rules out the 4-linear forms built from 4
fundamentals, as a possible source of deformation in $\mathcal{N}=8$
supergravity.

%%%%%%%%%%%%%%%%%%%%%%%%%%%%%%%%%%%%%%%%%%%%%%%%%%%%%%%%%%%%%%%%%%%%%%
%%%%%%%%%%%%%%%%%%%%%%%%%%%%%%%%%%%%%%%%%%%%%%%%%%%%%%%%%%%%%%%%%%%%%%
%%%%%%%%%%%%%%%%%%%%%%%%%%%%%%%%%%%%%%%%%%%%%%%%%%%%%%%%%%%%%%%%%%%%%%
%%%%%%%%%%%%%%%%%%%%%%%%%%%%%%%%%%%%%%%%%%%%%%%%%%%%%%%%%%%%%%%%%%%%%%
%%%%%%%%%%%%%%%%%%%%%%%%%%%%%%%%%%%%%%%%%%%%%%%%%%%%%%%%%%%%%%%%%%%%%%

\section{Reduction to smaller, non-degenerate and degenerate groups of type
  $E7$}
\label{sec-reduction}

Here we would like to study the cases with smaller duality symmetry group, for
example the case of $\mathcal{N}=4$ supergravity with $SL(2,\mathbb{R})\times
SU(4)$ duality, representing the non-degenerate group of type $E7$, and the
case of Born Infeld model with derivatives with $U(1)$ duality, without
scalars, representing the degenerate case of group of type $E7$.

%%%%%%%%%%%%%%%%%%%%%%%%%%%%%%%%%%%%%%%%%%%%%%%%%%%%%%%%%%%%%%%%%%%%%%
%%%%%%%%%%%%%%%%%%%%%%%%%%%%%%%%%%%%%%%%%%%%%%%%%%%%%%%%%%%%%%%%%%%%%%
%%%%%%%%%%%%%%%%%%%%%%%%%%%%%%%%%%%%%%%%%%%%%%%%%%%%%%%%%%%%%%%%%%%%%%
%%%%%%%%%%%%%%%%%%%%%%%%%%%%%%%%%%%%%%%%%%%%%%%%%%%%%%%%%%%%%%%%%%%%%%
%%%%%%%%%%%%%%%%%%%%%%%%%%%%%%%%%%%%%%%%%%%%%%%%%%%%%%%%%%%%%%%%%%%%%%

\subsection{Reduction to $SL(2,\mathbb{R})\times SU(4)$}

In our first example we consider the reduction of the $E_{7(7)}$\, duality
group of $\mathcal{N}=8$ supergravity to the $SL(2,\mathbb{R})\times SU(4)$
duality group of pure $\mathcal{N}=4$ supergravity. This duality group is of
type $E7$, a class defined by Brown in \cite{Brown} and studied in the context
of extended supergravities in \cite{Ferrara:2011dz}. The reason for our
interest in $\mathcal{N}=4$ supergravity is the recent computation of the
3-loop four-particle amplitude in this theory, which was found to be
divergence-free in \cite{Bern:2012cd}.

To get pure $\mathcal{N}=4$ supergravity from $\mathcal{N}=8$ we restrict
ourselves to the $SU(4)\times SL(2,\mathbb{R})$ subgroup of $E_{7(7)}$.  By
restricting the range of the indices in the ${\bf 56}$ of $E_{7(7)}$\
$(F^{ij},G_{ij})$ to $i, j=1,\cdots , 4$ we get the $(\mathbf{6}, \mathbf{
  2})$ of $SO(6)\times SL(2,\mathbb{R})$: $F^{ij}$ and $G_{ij}$ as $SO(6)\sim
SU(4)$ antisymmetric tensors and the pair $(F,G)$ transforms as an
$SL(2,\mathbb{R})$ doublet. We can also use a complex bassis
\begin{equation}
\label{4of56}
\overline{\mathcal{F}}^{\, AB}(p_{I}) \equiv 
\tfrac{1}{4\sqrt{2}}\left(F^{ij}(p_{I})-iG_{ij}(p_{I}) \right)
\Gamma^{ij}{}_{AB}\, ,
\end{equation}
where now $A,B=1,\cdots,4$. In this basis, the $SU(4)\times SL(2,\mathbb{R})$
invariant is given by Eqs.~(\ref{J}),(\ref{O}) and (\ref{inv}).  

With the same restriction in the range of the indices, the rest of the
formulae discussed in the previous sections, and in particular the invariants
$J_{4}^{\prime}$ in (\ref{eq:J4prime}) and $\diamondsuit^{\prime}$ in
(\ref{eq:diamondprime}), reduce to equivalent formulae for the $\mathcal{N}=4$
case.  The Pfaffian terms in the $\mathcal{N}=8$ invariants (\ref{eq:J4prime})
and (\ref{eq:diamondprime}) drop since in $\mathcal{N}=4$ model the totally
antisymmetric 8-rank tensor required in Eq.~(\ref{Pf}) is absent. However, the
trace of four operators in the first and second line and the square of the
trace of two operators in the third line (both in (\ref{eq:J4prime}) and
(\ref{eq:diamondprime})) remain independent. This is a property of the groups
of type $E7$ \cite{Brown,Ferrara:2011dz}. The quartic invariant is not a
square of the quadratic one since the symmetric bilinear form does not
exist\footnote{In $d=6$ maximal supergravity there is a 3-loop divergence
  \cite{Bern:2008pv}. This requires a separate detailed investigation in the
  context of the $SO(5,5)$ duality symmetry of this theory. However, as a
  comment to this investigation of $d=4$ extended supergravities, we would
  like to point out that $SO(5,5)$ is not a group of type $E7$. It has a
  symmetric bilinear form since there is a constant $SO(5,5)$ metric, as in
  any orthogonal group. This suggests that the extension of the comparison of
  duality invariants with tensor field amplitudes made in this paper to the
  $d=6$ case may yield different results, in line with other models of
  degenerate groups of type $E7$, when the restoration of duality is possible
  via the deformation procedure of \cite{Bossard:2011ij,Carrasco:2011jv}.
}. Therefore, the term with the trace of four operators needs help from the
term with the square of the trace of two operators, to form an $SU(4)\times
SL(2,\mathbb{R})$ invariant. Each of these terms is separately $SU(4)$
invariant, but only together, in the combination shown in
Eqs.~(\ref{eq:J4prime}) or (\ref{eq:diamondprime}), they are $SU(4)\times
SL(2,\mathbb{R})$ invariant.
 
Having constructed the new $SU(4)\times SL(2,\mathbb{R})$ invariant relevant
for amplitudes, we may now proceed with the same procedure we used for
$\mathcal{N}=8$ amplitudes and described in the previous sections. It is easy
to see that all the steps leading to the conclusion that the amplitude
disagrees with the invariant in the $\mathcal{N}=8$ case remains valid for
$\mathcal{N}=4$ supergravity.

%%%%%%%%%%%%%%%%%%%%%%%%%%%%%%%%%%%%%%%%%%%%%%%%%%%%%%%%%%%%%%%%%%%%%%
%%%%%%%%%%%%%%%%%%%%%%%%%%%%%%%%%%%%%%%%%%%%%%%%%%%%%%%%%%%%%%%%%%%%%%
%%%%%%%%%%%%%%%%%%%%%%%%%%%%%%%%%%%%%%%%%%%%%%%%%%%%%%%%%%%%%%%%%%%%%%
%%%%%%%%%%%%%%%%%%%%%%%%%%%%%%%%%%%%%%%%%%%%%%%%%%%%%%%%%%%%%%%%%%%%%%
%%%%%%%%%%%%%%%%%%%%%%%%%%%%%%%%%%%%%%%%%%%%%%%%%%%%%%%%%%%%%%%%%%%%%%

\subsection{Reduction to $U(1)$ with no scalars}

The source of deformation used in \cite{Chemissany:2011yv} for the case of the
open string or the D3 brane 1-loop corrections \cite{Shmakova:1999ai},
depends\footnote{In \cite{Chemissany:2011yv} $\overline{T}$ is denoted by
  $T^{*}$.}  on $T= F-iG$ and $\overline{T}=F+i G$ so that
\begin{equation}
\label{manifest4dF}
\begin{array}{rcl}
\mathcal{I}
& = &
{\displaystyle\frac{\lambda}{2^{4}}}\,
t^{(8)}_{abcd}
\left[\, 
2\partial_{\mu}\overline{T}^{+\,a}\partial^{\mu}T^{-\,b}
\partial_{\nu}\overline{T}^{+\,c}\partial^{\nu}T^{-\,d}
+ 
\partial_{\mu}\overline{T}^{+\,a}\partial_{\nu} T^{-\,b}
\partial^{\mu}\overline{T}^{+\,c}\partial^{\nu}T^{-\,d}
\,
\right]
\\
& & \\
& = & 
{\displaystyle\frac{\lambda}{2^{4}}}\,
t^{(8)}_{abcd}
\left[ \,
\partial_{\mu}\overline{T}^{+\,a}\partial^{\mu}T^{-\,b}
\partial_{\nu}\overline{T}^{+\,c}\partial^{\nu}T^{-\,d}
+ 
\partial_{\mu}\overline{T}^{+\,a}\partial_{\nu} T^{-\,b}
\partial^{\mu}\overline{T}^{+\,c}\partial^{\nu}T^{-\,d}
\right.
\\
& & \\
& & 
\left.
+
\partial_{\mu}\overline{T}^{+\,a}\partial_{\nu}T^{-\,b}
\partial^{\nu}\overline{T}^{+\,c}\partial^{\mu}T^{-\,d}
\,
\right]\, ,
\end{array}
\end{equation}
where we have used the full symmetry of $t^{(8)}$ in the $a,b,c,d$ indices.
This expression corresponds to
\begin{equation}
\label{manifest4dFmom}
\mathcal{I}=
\frac{\lambda}{2^{6}}\,
t^{(8)}_{abcd} 
\left[ \
\overline{T}^{+\,a}(p_{1}) T^{-\,b}(p_{2}) 
\overline{T}^{+\,c}(p_{3})T^{-\,d}(p_{4})
\right]  (s^{2}+t^{2} + u^{2}) \, ,
\end{equation}
in momentum space.

If we start with the $E_{7(7)}$\, invariant 
\begin{equation}
\frac{\lambda}{2^{2}}\, t^{(8)}\cdot \diamondsuit_{(8)}
\left(
\partial_{\mu}\mathcal{F},\partial_{\nu}\mathcal{F},
\partial^{\mu}\mathcal{F},\partial^{\nu}\mathcal{F}
 \right)\, ,
\end{equation}
and reduce it to $U(1)$, the Pfaffians vanish and all the traces become
ordinary products and the invariant reduces to exactly (\ref{manifest4dF})
after identification of $\mathcal{F}$ with $T$, which is the manifestly
$U(1)$-invariant source of deformation used in \cite{Chemissany:2011yv}.
Using the linear selfduality constraint
\begin{equation}
T^{+}=0\, ,
\end{equation}
the expression collapses into
\begin{equation}
\label{S1susy}
S^{(1)}= 
\frac{\lambda}{2^{4}}\,  \int d^{4}x\, t^{(8)}_{abcd}
\partial_{\mu}F^{a}\partial^{\mu}F^{b}\partial^{\nu}F^{c}\partial_{\nu}F^{d}\, .
\end{equation} 
This confirms that the source of deformation when linear duality constraint is
imposed, reduces to the correct quantum correction to the 4-vector amplitude
from open string or D3 branes.

In the $U(1)$ case the amplitude has no internal indices, which brings the
expression to the following one
\begin{equation}
\label{4vectorsU1}
M_{4vec}( b^{-}, b^{-}, b_{+}, b_{+}) 
= \lambda \langle12\rangle^{2}[34]^{2}
\left[\, 
s\, u\, 
+
s\, t \, 
+
t\, u\, 
\right]\, .
\end{equation}
This expression is in agreement with the one in \cite{Chemissany:2011yv},
(compare with Eq.~(4.2) there).

%%%%%%%%%%%%%%%%%%%%%%%%%%%%%%%%%%%%%%%%%%%%%%%%%%%%%%%%%%%%%%%%%%%%%%
%%%%%%%%%%%%%%%%%%%%%%%%%%%%%%%%%%%%%%%%%%%%%%%%%%%%%%%%%%%%%%%%%%%%%%
%%%%%%%%%%%%%%%%%%%%%%%%%%%%%%%%%%%%%%%%%%%%%%%%%%%%%%%%%%%%%%%%%%%%%%
%%%%%%%%%%%%%%%%%%%%%%%%%%%%%%%%%%%%%%%%%%%%%%%%%%%%%%%%%%%%%%%%%%%%%%
%%%%%%%%%%%%%%%%%%%%%%%%%%%%%%%%%%%%%%%%%%%%%%%%%%%%%%%%%%%%%%%%%%%%%%

\section{Discussion}
\label{sec-discussion}

Here we will discuss shortly the two prototypes of relations between UV
divergences and duality\footnote{Recently the issue was raised in
  \cite{Roiban:2012gi} as to whether duality symmetry has to control the
  quantum theory.} which came to light as a consequence of the studies
performed in this paper. The first one has to do with computations in
\cite{Shmakova:1999ai} and \cite{DeGiovanni:1999hr} where certain UV
divergences were found, in agreement with duality symmetry. The second one has
to do with computations in \cite{Bern:2007hh} and in \cite{Bern:2012cd} where
the UV divergences were not found, in agreement with duality current
conservation.
  
As an example of the first prototype consider the computation of the one-loop
corrections to the $\kappa$-symmetric D3 brane action performed in
\cite{Shmakova:1999ai}. The action consists of 2 terms, the Born-Infeld one
and the Wess-Zumino one. After gauge-fixing the local $\kappa$-symmetry the
bosonic part of action, up to 4th order in fields is given by (Eq.~(37) in
\cite{Shmakova:1999ai}.
\begin{equation}
  L^{4}_{\rm BI+WZ}
  =
  -\tfrac{1}{4} F_{\mu\nu} F^{\mu\nu}
  +\frac{\alpha^{2}}{8} 
  (F_{\mu\nu} F^{\nu\lambda} F_{\lambda \delta} F^{\delta\mu}
  - \tfrac{1}{4} (F_{\mu\nu} F^{\mu\nu})^{2}) + \ldots
\end{equation}
where the terms indicated by the dots depend on 6 scalars, vectors and
fermions. Note that the term quartic in vectors is
\begin{equation}
F_{\mu\nu} F^{\nu\lambda} F_{\lambda \delta} F^{\delta \mu}
- 
\tfrac{1}{4} (F_{\mu\nu} F^{\mu\nu})^{2}\equiv F^{4}
- 
\tfrac{1}{4} (F^{2})^{2} 
= 
\tfrac{1}{4!}
t^{(8)}_{\mu_{1}\nu_{1}\mu_{2}\nu_{2}\mu_{3}\nu_{3}\mu_{4}\nu_{4}} 
F^{\mu_{1} \nu_{1}} F^{\mu_{2} \nu_{2}} F^{\mu_{3} \nu_{3}} F^{\mu_{4} \nu_{4}}.
\end{equation}

The 1-loop UV divergence in $d=4$ Feynman graphs was computed in
\cite{Shmakova:1999ai}, with the result
\begin{equation}
\label{Marina}
- \frac{\alpha^{4}}{(4\pi)^{2}} 
\frac{1}{16 \, \epsilon}
(s^{2}+t^{2}+u^{2})
t^{(8)}_{\mu_{1}\nu_{1}\mu_{2}\nu_{2}\mu_{3}\nu_{3}\mu_{4}\nu_{4}} 
F^{\mu_{1} \nu_{1}}(p_{1}) F^{\mu_{2} \nu_{2}}(p_{2}) 
F^{\mu_{3}\nu_{3}}(p_{3}) F^{\mu_{4} \nu_{4}}(p_{4}).
\end{equation}

The fact that the expected UV divergence does show up in this computation may
be interpreted here as follows:

\begin{quote} {\it Whenever the candidate counterterms may violate the NGZ
    duality current conservation, but the complete deformation procedure of
    \cite{Bossard:2011ij,Carrasco:2011jv} adding higher order terms, restoring
    duality, can be performed, the expected candidate counterterm is not
    forbidden and there is a UV divergence}.
\end{quote}

The particular case of the Born-Infeld model with higher derivatives and
$U(1)$ duality was worked out in \cite{Chemissany:2011yv} and the existence of
the procedure of reconstructing the Noether-Gaillard-Zumino current
conservation was demonstrated when higher order terms were recursively
produced.

In a related computation performed in \cite{DeGiovanni:1999hr} one starts with
a manifestly $\mathcal{N}=2$ supersymmetric $d=4$ BI action and computes the
1-loop UV divergence using the quantization in terms of $\mathcal{N}=1$
superfields. The Feynman supergraphs computation yields the expected UV
divergence which takes the manifestly $\mathcal{N}=1$ supersymmetric form
\begin{equation}
\label{Daniela}
\frac{\alpha^{4}}{(4\pi)^{2}}
\frac{1}{8 \epsilon}(s^{2}+ \tfrac{4}{3} t^{2} ) W^{2}(1,2)
\overline{W}^{2}(3,4)\, .
\end{equation}
Its bosonic part, due to the complete symmetry of $F^{4}- \tfrac{1}{4}
(F^{2})^{2}$ in $s,t,u$ variables, is shown to be proportional to an
expression given in (\ref{Marina}). The existence of a 1-loop UV divergence in
this case, again, is supported by the construction of $U(1)$ duality invariant
manifestly supersymmetric actions in \cite{Broedel:2012gf} where the first
quartic term would violate the duality current conservation but higher-order
terms produced algorithmically restore the Noether-Gaillard-Zumino identity.

From this perspective our work suggests the following possible interpretation
of the fact that $\mathcal{N}=8$ and $\mathcal{N}=4$ supergravities are UV
finite at the 3-loop order:

\begin{quote} {\it Whenever the candidate counterterms may violate the
    duality current conservation, and there is an obstruction for the complete
    procedure adding higher order terms restoring the Noether-Gaillard-Zumino
    current conservation, the expected candidate counterterm is forbidden and
    there is no UV divergence}.
\end{quote}

In this paper we have produced evidence that in the supergravities whose
duality group is of type $E7$, with $\mathcal{N}=8$ and $\mathcal{N}=4$ as
explicit examples, the deformation procedure of
\cite{Bossard:2011ij,Carrasco:2011jv,Chemissany:2011yv,Broedel:2012gf}
required for the restoration of the duality current conservation seems to
encounter an obstruction: the existence of a superinvariant initial source of
deformation depending on unconstrained doublets of electric and magnetic field
strengths, which is the cornerstone of the whole procedure, seems to be
incompatible with supersymmetry.
  
The evidence consists of two separate independent parts.

First, we have used the superspace construction of
\cite{Brink:1979nt,Howe:1981gz} which provides the candidate counterterms and
we have tried to promote them to initial sources of deformation. We have found
that relaxing the linear twisted self-duality in superspace violates the
integrability condition for the existence of the superspace 56-bein. In such
case, the proof that the graviphoton superfield is manifestly $E_{7(7)}$,
invariant becomes invalidated.

Secondly, we have used the fact that $E_{7(7)}$\, is based on exceptional
Jordan algebra $J_{3} ^{\mathbb{O}_S }$ over the split octonions
$\mathbb{O}_S$ \cite{Gunaydin:2000xr,Gunaydin:2009zza} to construct new
$E_{7(7)}$\, invariants relevant to amplitudes and counterterms.  When we
compare the new invariants with the candidate divergent supergravity
amplitudes, we find an inconsistency: the dependence on Mandelstam variables
interferes with the internal structure of a generalized $E_{7(7)}$\, quartic
invariants which we have constructed. Namely, the trace of four operators and
square of the trace of two operators inside the invariant acquire different
dependences on $s,t,u$ in the 4-vector amplitude. This breaks $E_{7(7)}$
invariance: there is no initial source of deformation with required
properties, no consistent deformation procedure and no UV divergence
prediction. This agrees with the 3- and 4-loop computations of
\cite{Bern:2007hh,Bern:2008pv} for $\mathcal{N}=8$.  The same argument works
for all groups of type $E7$, as long as they are not degenerate, which
includes the duality group of the pure $\mathcal{N}=4$ theory. This leads to a
prediction that there is no UV divergence, in agreement with the the 3-loop
computations in \cite{Bern:2012cd}.

In the reduction to $U(1)$ the new $E_{7(7)}$\, invariants reduce to the ones
considered in the Born-Infeld models with higher derivatives of
\cite{Chemissany:2011yv}. In that case there is no difference between the
internal symmetry trace of four operators and the square of the trace of two
operators and the functions of the Mandelstam variables do not introduce any
discrepancy. There is a manifest $U(1)$ duality source of deformation with the
required properties, there is a consistent deformation procedure and there are
UV divergences, as shown in \cite{Shmakova:1999ai} and
\cite{DeGiovanni:1999hr}.

In the future, hopefully, there will be more computational examples and
theoretical studies of the role of duality symmetries in quantum theories,
like \cite{Roiban:2012gi,Padua}, to test the ideas of this paper.

%%%%%%%%%%%%%%%%%%%%%%%%%%%%%%%%%%%%%%%%%%%%%%%%%%%%%%%%%%%%%%%%%%%%%%
%%%%%%%%%%%%%%%%%%%%%%%%%%%%%%%%%%%%%%%%%%%%%%%%%%%%%%%%%%%%%%%%%%%%%%
%%%%%%%%%%%%%%%%%%%%%%%%%%%%%%%%%%%%%%%%%%%%%%%%%%%%%%%%%%%%%%%%%%%%%%
%%%%%%%%%%%%%%%%%%%%%%%%%%%%%%%%%%%%%%%%%%%%%%%%%%%%%%%%%%%%%%%%%%%%%%

\section*{Acknowledgments}

We are most grateful to our collaborators on related projects, J.~Broedel,
J.J.~Carrasco, W.~Chemissany, S.~Ferrara, R.~Roiban.  We had valuable
discussions with G.~Bossard, D.~Freedman, M.~G\"unaydin, S.~Ketov, S.~Kuzenko,
H.~Nicolai, D.~Sorokin, S.~Theisen, M.~Tonin and A.~Tseytlin.  The work of
R.K.~was supported by SITP and NSF grant PHY-0756174.  The work of T.O.~has
been supported in part by the Spanish Ministry of Science and Education grant
FPA2009-07692, the Comunidad de Madrid grant HEPHACOS S2009ESP-1473, and the
Spanish Con\-solider-Ingenio 2010 program CPAN CSD2007-00042.  T.O.~is
grateful to SITP for the hospitality, extended to him at Stanford and R.K.~is
grateful to IFT for the hospitality, extended to her in Madrid.  T.O.~wishes
to thank M.M.~Fern\'andez for her unfaltering support.

%%%%%%%%%%%%%%%%%%%%%%%%%%%%%%%%%%%%%%%%%%%%%%%%%%%%%%%%%%%%%%%%%%%%%%
%%%%%%%%%%%%%%%%%%%%%%%%%%%%%%%%%%%%%%%%%%%%%%%%%%%%%%%%%%%%%%%%%%%%%%
%%%%%%%%%%%%%%%%%%%%%%%%%%%%%%%%%%%%%%%%%%%%%%%%%%%%%%%%%%%%%%%%%%%%%%
%%%%%%%%%%%%%%%%%%%%%%%%%%%%%%%%%%%%%%%%%%%%%%%%%%%%%%%%%%%%%%%%%%%%%%
\appendix
%%%%%%%%%%%%%%%%%%%%%%%%%%%%%%%%%%%%%%%%%%%%%%%%%%%%%%%%%%%%%%%%%%%%%%
%%%%%%%%%%%%%%%%%%%%%%%%%%%%%%%%%%%%%%%%%%%%%%%%%%%%%%%%%%%%%%%%%%%%%%
%%%%%%%%%%%%%%%%%%%%%%%%%%%%%%%%%%%%%%%%%%%%%%%%%%%%%%%%%%%%%%%%%%%%%%
%%%%%%%%%%%%%%%%%%%%%%%%%%%%%%%%%%%%%%%%%%%%%%%%%%%%%%%%%%%%%%%%%%%%%%

\section{On $\mathcal{N}=8$ superspace, counterterms 
and candidates for  initial source of deformation}
\label{ap-superspace}

The on-shell superspace of $\mathcal{N}=8$ supergravity was constructed in
\cite{Brink:1979nt,Howe:1981gz}.  The non-linear, geometric invariant
counterterms, starting from the 8-loop order, were proposed in
\cite{Kallosh:1980fi}. For example, for the 8-loop order one has
\begin{equation}
S^{8} \sim \kappa^{14} \int d^{4} x \; d^{32} \theta  \; {\rm Ber} \, E \; 
\chi_{IJK \alpha} (x, \theta)  \overline \chi^{IJK \dot{\alpha}} (x, \theta)  
\chi_{MNL}{}^{\alpha} (x, \theta) \overline \chi^{MNL}{}_{\dot{\alpha}}(x, \theta)  \ .
\label{S8}
\end{equation}
Here 
\begin{equation}
\label{Tor} 
T_{I \dot{\alpha} \,  J \dot{\beta} , \, K \alpha}(x, \theta) 
= 
\epsilon _{\dot{\alpha} \dot{\beta}} \chi_{IJK \alpha} (x, \theta)
\end{equation} 
is the superspace torsion superfield whose first component is a spinor field
$\chi_{IJK\, \alpha}(x)$. The existence of a generic $L$-loop counterterm
where the superspace has a well-defined $SU(8)$- and Lorentz-covariant
derivatives $D_{\alpha}^{I}$, $\overline{D}_{\dot{\alpha} I}$ and $D_{\alpha
  \dot{\alpha}}$, is based on the fact that the superspace construction
provides a solution of geometric Bianchi identities. Therefore one can
increase the dimension of the counterterm easily by inserting covariant
derivatives and using more torsions and curvatures in the superinvariants
analogous to (\ref{S8}).

To promote this construction to the level of the source of deformation
\cite{Bossard:2011ij,Carrasco:2011jv} of the linear twisted self-duality
constraint, valid in classical theory, we have to relax the linear twisted
self-duality constraint. This requires to clarify some subtleties in the
superspace construction \cite{Brink:1979nt,Howe:1981gz}, which were not
important in the past.

The basic variables are the (super-) vielbein $E_{M}{}^{A}$ and the $SL(2,
\mathbb{C}) \times SU(8)$ connection $\Omega_{MA}{}^{B}$ from which the
torsion $T_{AB}{}^{C}$ and the curvature $R_{ABC}{}^{D}$ are constructed in
the usual way and satisfy, from their definitions, the following Bianchi
identities:
\begin{equation}
DT^{A}= E^{B} \wedge R_{B}{}^{A}, \qquad DR_{B}{}^{A}=0
\label{Bian}
\end{equation}
Notations and conventions are the same as in \cite{Howe:1981gz} except for the
$SU(8)$ indices that we denote by capital $I,J,\ldots$ instead of lowercase
$i,j,\ldots$. In particular, the tangent space rotation is
\begin{equation}
\delta E^{A}= E^{B} L_{A}{}^{B}, 
\label{tan}
\end{equation}
where the matrix $L_{A}{}^{B}$ is
\begin{equation}
\label{L}
L_{ab}=- L_{ba}\, ,\,\,\, 
L_{\alpha \beta} =\tfrac{1}{2} (\sigma^{ab})_{\alpha \beta}  L_{ab}\, ,\,\,\,  
L_{\underline{\alpha}}{}^{\underline{\beta}}
= 
\delta^{J}{}_{I} L_{ \alpha}{}^{ \beta} +\delta_{ \alpha}{}^{ \beta}
L^{J}{}_{I}\, ,\,\,\, 
\overline{L}_{I}{}^{J}= - L^{J}{}_{I}\, ,\,\,\,
L_{\underline {\dot{\alpha}}}{}^{\underline {\dot{\beta}}} 
=
- (\overline {L_{\underline{\alpha}}{}^{\underline{\beta}}})\, .
\end{equation}

The summary of this part in Section~4 of \cite{Howe:1981gz} is that a complete
solution to the geometrical Bianchi identities can be given in terms of the
following set of covariant superfields:
\begin{equation}
\chi_{\alpha [IJK]}\, , 
\quad 
S^{(IJ)}\, , 
\quad 
N_{(\alpha\beta)}^{[IJ]}\, , 
\quad 
G_{\alpha \dot{\beta} J}{}^{I}\, , 
\quad 
H_{\alpha \dot{\beta} J}^{I}\, .
\end{equation}
These superfields appear in the superspace torsion as follows, in addition to
(\ref{Tor})
\begin{eqnarray}
T_{\alpha \dot{\alpha},\,}{}_{\beta,\,}^{J,\,}{}_{\dot{\gamma}}^{K}
& = & 
i(\epsilon_{\alpha \beta} \overline{M}_{\dot{\alpha} \dot \gamma}^{JK} 
+ 
\epsilon _{\dot{\alpha} \dot{\gamma}} N_{\alpha \beta}^{JK} 
+ 
\epsilon_{\alpha \beta} \epsilon _{\dot{\alpha} \dot{\gamma}} S^{JK})\, ,
\\
& & \nonumber \\
T_{\alpha \dot{\alpha},\,}{}_{\beta,\,}^{J\,}{}_{\gamma  K}
& = & 
- i(\epsilon_{\alpha \gamma} G_{\beta}^{J}{}_{\dot{\alpha} K} 
+ 
\epsilon_{\beta\gamma} G^{J}_{\alpha \dot{\alpha} K} 
+  
\epsilon_{\alpha\beta} H^{J}_{\gamma \dot{\alpha} K})\, ,
\end{eqnarray}    
where 
\begin{equation}
M_{\alpha \beta IJ}= \tfrac{1}{6} D^{K}_{(\alpha} \chi_{\beta)IJK}  
\end{equation} 
and 
\begin{equation}
T^{I}_{\alpha \dot{\beta} J}{}^{c}
= 
-i \delta^{I}{}_{J} \sigma^{c}{}_{\alpha \dot{\beta}}\, .
\end{equation}

The components of the super curvature satisfy the Bianchi identities.

To identify the scalars and vectors in $\mathcal{N}=8$ superspace there are two options. 
\begin{enumerate}
\item In \cite{Brink:1979nt} it was suggested to use the ``somewhat specious''
  Bianchi identities as they have additional terms over and above the normal
  ones. The goal was to make a local $U(1)$ compatible with local $SU(8)$ and
  global $E_{7(7)}$ and at this point they introduced a ``non-geometric''
  1-form $P$ and a 2-form $F$ in their Table~4 on p.~271, so that in the end
  they provide a 56-bein in superspace.
  
  This reminds the 11-dimensional superspace construction
  \cite{Cremmer:1980ru} where in addition to geometric torsions and curvatures
  a 3-form gauge superfield $A$ and the 4-form $F=dA$ supercovariant field
  strength were introduced, and in the end all component structure of
  11-dimensional on-shell supergravity was presented in a superspace via a
  single superfield.
  
\item The second option, explained in Section~8 of \cite{Howe:1981gz} is to
  introduce additional 28 complex bosonic coordinates and assume that
  superfields do not depend on these new coordinates. However, the existence
  of these coordinates provides new components of torsion and
  curvatures. This, in turn, gives a geometric interpretation to the 1-form
  $P$ and the 2-form $F$ and to the identities they have to satisfy. In this
  second option the identities for the 1-form $P$ and the 2-form $F$ become
  the standard torsion-curvature Bianchi identities in this extended space.
  Since this version leads to the same results, we will only discuss the first
  one.

\end{enumerate}

Thus, we start with the end of Section~4 of \cite{Howe:1981gz} where only the
geometric Bianchi identities were solved. This leaves us with a solution for
the torsions and curvatures expressed via the unconstrained superfields
$\chi_{[IJK]\alpha }, S^{(IJ)},N_{(\alpha\beta)}^{[IJ]}$, $G^{I}_{\alpha
  \dot{\beta} J} $, $H^{I}_{\alpha \dot{\beta} J}$.

At this point the theory is not on shell (it is possible to continue with
off-shell conformal $\mathcal{N}\leq 4$ supergravity or on shell
Poincar\'e). However, in $\mathcal{N}=8$ case the scalar fields are not
present and $M_{\alpha \beta IJ}, \overline{M}_{\dot{\alpha} \dot{\beta}}^{IJ}$ is
not yet the field strength for a spin-1 field.  We now move on to Section~8 of
\cite{Howe:1981gz} to explain the origin of scalars and vectors in the
superspace.

We identify the scalar curvature in $SU(8)$ direction 
\begin{equation}
R^{IJ}{}_{KL}= 2 \delta^{[I}{}_{[K} R^{J]}{}_{L]}\, ,
\label{R0}\end{equation}
where $R^{I}{}_{K}$ is one of the components of the $SL(2, \mathbb{C}) \times
SU(8)$ curvature defined in the tangent space in Eqs.~(\ref{tan}), (\ref{L}) and
determined in terms of superfields $\chi_{[IJK]\alpha }, \, S^{(IJ)}, \,
N_{(\alpha\beta)}^{[IJ]}, \, G^{I}_{\alpha \dot{\beta} J} , \, H^{I}_{\alpha
  \dot{\beta} J}$ in \cite{Howe:1981gz}.

We now additionally require that 
\begin{equation}
R^{IJ}{}_{KL}= - \overline{P}^{IJMN}\wedge P_{KLMN}
\label{R}\end{equation}
where the 1-form $P_{IJKL}$ is a new object not present in geometry such that
\begin{equation}
\label {DP}
DP_{IJKL}=0\, .
\end{equation}

These two equations above serve the following purpose: (\ref{R}), (\ref{DP})
are integrability conditions for the existence of a 56-bein in a superspace. A
 1-form $\hat{\Omega}$ with the properties
\begin{equation}
\hat{\Omega} = 
\left(
\begin{array}{cc}
\Omega^{IJ}{}_{KL} & - \overline{P}^{IJKL} \\
& \\
- P_{IJKL} & \overline{\Omega}_{IJ}{}^{KL} \\
\end{array}
\right) \, , 
\qquad 
d\hat{\Omega} + \hat{\Omega}\wedge \hat{\Omega} =0\, ,
\label{flat}
\end{equation}
is postulated to exist. 
Here the connection 1-form is $ \Omega^{IJ}{}_{KL}= 2 \delta ^{[I}{}_{[K}
\Omega^{J]}{}_{L]}$. If Eq.~(\ref{flat}) is valid, the existence of a
$56\times 56$-dimensional matrix ${\cal S}\in E_{7(7)}$ 
such that 
\begin{equation}
\hat{\Omega} = - {\cal S}^{-1} d {\cal S}\, ,
\label{V}
\end{equation}
is guaranteed.

$\hat{\Omega}$ is a superspace vielbein, a bridge between the local $SU(8)$
and the global $E_{7(7)}$. The scalar fields are described by  ${\cal S}$.

Then, a 2-form $F_{IJ},\overline{F}^{KL}$ not present in the geometry and such that
\begin{equation}
D F_{IJ}= \overline{F}^{KL} \wedge P_{KLIJ}\, ,
\label{DF}
\end{equation}
is introduced.  The choice to solve this one is \cite{Howe:1981gz}
\begin{eqnarray}
F^{IJ}_{\alpha \beta , KL} 
& = & 
-2 i \epsilon_{\alpha \beta} \delta ^{IJ}{}_{[KL]}\, ,
\\
& & \nonumber \\
F_{\alpha \dot{\alpha} , \dot{\beta} J, KL}
& = &
 \epsilon _{\dot{\alpha} \dot{\beta}} \chi_{JKL\alpha }\, ,
\\
& & \nonumber \\
F_{\alpha \dot{\alpha}, \beta \dot{\beta}, IJ}
& = &
 -i \epsilon_{\dot{\alpha} \dot{\beta}} M_{\alpha \beta IJ}
-i \epsilon_{ \alpha  \beta} \overline{N}_{\dot{\alpha} \dot{\beta} IJ}\, ,
\end{eqnarray}
and  conjugate, the rest vanishes. For example, 
\begin{equation}
\overline{F}_{\alpha \dot{\alpha}, \beta \dot{\beta}}^{IJ}
= 
i \epsilon_{\alpha  \beta} \overline{M}_{\dot{\alpha} \dot{\beta}}^{IJ}
+i \epsilon_{\dot  \alpha  \dot{\beta}}  N_{\alpha  \beta}^{IJ}\, .
\label{F}
\end{equation}

To solve (\ref{DP}) and (\ref{DF}) one finds according to \cite{Howe:1981gz} that
\begin{equation}
\label{chi}
\begin{array}{rcl}
P^{I}_{\alpha JKLM}
& = & 
2 \delta ^{I}{}_{[J} \chi_{KLM] \alpha }\, ,
\\
& & \\
P_{\alpha \dot{\alpha} IJKL}
& = &
-\tfrac{1}{2} \overline{D}_{\dot{\alpha} [I} \chi _{JKL] \alpha}\, ,
\end{array}
\end{equation}
and
\begin{equation}
  \begin{array}{rcl}
D_{(\alpha}^{I} \chi^{JKLMN}{}_{\beta)}
& = & 
M_{\alpha\beta}{}^{IJKLMN}\, ,
\\ 
& & \\
D_{\alpha}^{I} \chi^{JKLMN \alpha }
& = &
-5 \overline{\chi}^{I[JK}{}_{\dot{\alpha}} \overline{\chi}^{LMN] \dot{\alpha}}\, ,
\end{array}
\end{equation}
where
\begin{equation}
  \begin{array}{rcl}
M_{\alpha\beta}{}^{IJKLMN}
& \equiv &
\epsilon^{IJKLMNPQ} M_{\alpha\beta PQ}\, ,
\\
& & \\
\chi^{IJKLM \alpha}
& \equiv &
\epsilon^{IJKLMNPQ}\chi_{NPQ}{}^{\alpha}\, .
\end{array}
\end{equation}

At this point $M_{\alpha\beta IJ}$ is identified as a self-dual $SU(8)$
covariant vector field strength and the anti-self-dual one,
$\overline{N}_{\dot{\alpha} \dot{\beta} IJ} $ is arbitrary. This is what we need
since the {\it linear twisted self-duality condition in absence of fermions}
requires that
\begin{equation}
\overline{N}_{\dot{\alpha} \dot{\beta} IJ}=N_{\alpha \beta}^{IJ}=0\, ,
\end{equation}
and we have 28 $M_{\alpha \beta IJ}$ and the conjugate 28 $\overline{M}_{\dot{\alpha}
  \dot{\beta}}^{IJ}$.  Relaxing the linear twisted self-duality condition in
absence of fermions simply means that $N_{\alpha \beta}^{IJ}$ and its
conjugate are unconstrained.

When there are fermions present, the linear twisted self-duality condition is
reduced to the requirement that the superfield $N_{\alpha \beta}^{IJ}$ is a
particular function of the fermion superfield $\chi_{IJK \alpha}$, as we will
see below.

%%%%%%%%%%%%%%%%%%%%%%%%%%%%%%%%%%%%%%%%%%%%%%%%%%%%%%%%%%%%%%%%%%%%%%
%%%%%%%%%%%%%%%%%%%%%%%%%%%%%%%%%%%%%%%%%%%%%%%%%%%%%%%%%%%%%%%%%%%%%%
%%%%%%%%%%%%%%%%%%%%%%%%%%%%%%%%%%%%%%%%%%%%%%%%%%%%%%%%%%%%%%%%%%%%%%
%%%%%%%%%%%%%%%%%%%%%%%%%%%%%%%%%%%%%%%%%%%%%%%%%%%%%%%%%%%%%%%%%%%%%%

\subsection{Consistency check}

So far, the superfield $\overline{N}_{\dot{\alpha} \dot{\beta} IJ}$ which appears
in the unconstrained $SU(8)$ vector field strength in Eq.~(\ref{F}) was not
required to satisfy any constraints. But we have not yet checked
Eq.~(\ref{R0}) which is equivalent to Eq.~(\ref{L}). It means that the
computation of the curvature $R^{I}{}_{J}$s done in the first (geometric) part
(before the scalars and vectors were introduced) in Section~4 of
\cite{Howe:1981gz}, has to be used. In Section~4 all geometric identities (in
a smaller space) were satisfied with and arbitrary, unconstrained,
$\overline{N}_{\dot{\alpha} \dot{\beta} IJ}$, and all the answers for $R^{I}{}_{J}$
are given. We have to extract the value of $R^{I}{}_{j}$ from there and plug
it into the integrability equation for the existence of the superspace 56-bein
Eq.~(\ref{R}) via Eq.~(\ref{R0}). This is Eq.~(8.17) of \cite{Howe:1981gz} and
it requires that $ R^{I}{}_{J}$ computed in Section~4 is compatible with
existence of the 1- and 2-forms so that Eq.~(\ref{R}) is satisfied.  But since
$ R^{I}{}_{J}$ depends on the unconstrained $\overline{N}_{\dot{\alpha} \dot{\beta}
  IJ}$ and it is part of an unconstrained vector field strength (\ref{F}), we
have a chance to see if it is consistent with identities to keep $\overline
N_{\dot{\alpha} \dot{\beta} IJ}$ unconstrained, i.e.~to relax the linear
twisted self-duality constraint.

\begin{description}
\item[The vector-vector part] given in Eq.~(4.42) of \cite{Howe:1981gz}
  \begin{equation}
    R_{\alpha \dot{\alpha}, \beta \dot{\beta}}{}^{K}{}_{L}
    = 
    \epsilon_{\dot{\alpha} \dot{\beta}} X_{\alpha \beta}{}^{K}{}_{L} 
    - \epsilon_{ \alpha  \beta} \overline{X}_{\dot{\alpha} \dot{\beta}}{}^{K}{}_{L}\, .
  \end{equation}
  The expression for $ X_{\alpha \beta}{}^{K}{}_{L}$ and its conjugate can be
  extracted, in principle, from \cite{Howe:1981gz}. It depends on the
  superfields $\chi, S, N, G, H$ and their covariant derivatives. This
  dependence is complicated and it is not clear if one can make it useful.
  For example, if we tried to find the value of the bosonic part of $X_{\alpha
    \beta}{}^{K}{}_{L}$ with $N$ unconstrained, we may or may not find a
  restriction on $N$.

  Meanwhile, in the bosonic case, one simply constructs the bosonic 56-bein
  from scalars, as in \cite{Cremmer:1979up,de Wit:1982ig}. Therefore, there is
  no need to provide the integrability conditions for its existence. But this
  procedure may not be the same as extracting the bosonic terms from the
  superspace. So, we leave this part for future studies and proceed to study
  other components of the curvature.

\item[The spinor-vector part] is in Eq.~(4.26) of \cite{Howe:1981gz} and,
  since it is fermionic, we cannot use the simple approach without
  fermions. Furthermore, it is rather messy with account of the gravitino
  superfield. So again, we leave this part unfinished and proceed to other
  study components of curvature.

\item[The spinor-conjugate spinor part] is given Eq.~(4.19) of
  \cite{Howe:1981gz}. But it only depends on $G, H$ and ot depend on $N$. It
  appears that both $G$ and $H$ have to vanish in absence of fermions.

\item[The spinor-spinor part] is given in Eq.~(4.14) of \cite{Howe:1981gz},
  it is concise and easy to analyze
  \begin{equation}
    R_{\alpha\beta,}^{I\, \, J,\, K}{}_{L}
    = 
    \epsilon_{\alpha \beta} (\delta^{I}{}_{L} S^{JK}
    - \delta^{J}{}_{L} S^{IK}
    - \overline{B}^{IJK}{}_{L})
    + \delta^{I}{}_{L} N^{JK}_{\alpha \beta}
    + \delta^{J}{}_{L} N^{IK}_{\alpha \beta}\, .
    \label{consistency}\end{equation}
  Here the superfields $S$ and $N$ are unconstrained, while, due to the
  geometric Bianchi identities of Section~4 of \cite{Howe:1981gz}
  $B_{IJK}{}^{L}= \tfrac{1}{2}D_{\alpha}^{L} \chi_{IJK}{}^{\alpha}$.  In
  absence of fermions, when $\chi_{IJK}{}^{\alpha}$ and $P^{IKMN}_{\alpha}
  \wedge P_{\beta JKLM}$ are not available, the r.h.s.~of
  $R_{\alpha\beta}^{I\, \, J,\, K}{}_{L}$ should vanish, according to
  Eq.~(\ref{curv}). Since each term in Eq.~(\ref{consistency}) has different
  symmetry properties, they all must vanish, namely {\it in absence of
    fermions}
  \begin{equation}
    S^{(JK)}=\overline{B}^{IJK}{}_{L}=N^{[JK]}_{\alpha \beta}=0\, .
  \end{equation}
  For the superfields the actual constraint is given by (\ref{consistency}),
  taken together with (\ref{curv}) in the form
  \begin{equation}
    R_{\alpha\beta,}^{I\, \, J,\, K}{}_{L}
    =  
    -\tfrac{1}{3\cdot 4!} \epsilon^{KMNPQRST} P^{I}_{\alpha QRST}\,  P^{J}_{\beta L
      MNP}\, ,
  \end{equation}
  with account of Eq.~(\ref{chi}).  This leads to the following constraints,
  \begin{eqnarray}
    \label{S}
    S^{IJ} & = & 0\, ,\\
    & & \nonumber \\
    \label{N2}
    N_{\alpha\beta}^{IJ}(x, \theta)
    & = &
    -\tfrac{1}{72} \epsilon^{IJKLMNPOQ}\chi_{KLM \alpha}\,\, \chi_{NPQ \beta} (x,
    \theta)\, .
  \end{eqnarray}

  The constraints (\ref{S}) and (\ref{N2}) in the on-shell superspace of
  \cite{Brink:1979nt} are imposed from the start to solve the Bianchi
  identities (\ref{Bian}) (see Table~2 there).  One may have thought that they
  are necessary for the existence of the one-shell conditions, with which we
  are not concerned here. However, we may start with the solution of Bianchi
  identities in Section~4 of \cite{Howe:1981gz} with unconstrained
  $N_{\alpha\beta}^{IJ}$, and we would like to keep it that way. However, if
  we do that we find a discrepancy with the integrability condition for the
  existence of the superspce 56-bein since (\ref{N2}) follows from (\ref{R0})
  and (\ref{R}) and does not admit any possibility to relax it.

\end{description}

In conclusion, the superspace vielbein does not exist if Eq.~(\ref{N2}) is not
satisfied.  It is not clear, therefore, how exactly to proceed with the
``deformation program'' for $\mathcal{N}=8$ supergravity since the known
counterterms cannot be automatically promoted to an initial source of
deformation ${\cal I}$: their manifest $E_{7(7)}$\, invariance, manifest local
supersymmetry and manifest local $SU(8)$ symmetry are only valid when the
undeformed constraint Eq.~(\ref{N2}) is imposed.

In the models \cite{Bossard:2011ij} --\cite{Broedel:2012gf} this was not an
issue.  With a bit of work it was possible to find initial sources of
deformation with all required symmetry properties, and then it was possible to
produce new models in an algorithmic way. For example, in
\cite{Broedel:2012gf} the manifest $\mathcal{N}=2$ supersymmetry was not
affected by the deformation of the action. In $\mathcal{N}=8$ supergravity it
appears that the supersymmetry rules are affected by deformation of the action
and to find an initial source of deformation simultaneously consistent with
(deformed) duality and (deformed) supersymmetry is a problem. Knowing how to
solve it, one may construct an $\mathcal{N}=8$ Born-Infeld-type
supergravity. However, the absence of a solution is an indication of the UV
finiteness of $\mathcal{N}=8$ supergravity.

%%%%%%%%%%%%%%%%%%%%%%%%%%%%%%%%%%%%%%%%%%%%%%%%%%%%%%%%%%%%%%%%%%%%%%
%%%%%%%%%%%%%%%%%%%%%%%%%%%%%%%%%%%%%%%%%%%%%%%%%%%%%%%%%%%%%%%%%%%%%%
%%%%%%%%%%%%%%%%%%%%%%%%%%%%%%%%%%%%%%%%%%%%%%%%%%%%%%%%%%%%%%%%%%%%%%
%%%%%%%%%%%%%%%%%%%%%%%%%%%%%%%%%%%%%%%%%%%%%%%%%%%%%%%%%%%%%%%%%%%%%%
%%%%%%%%%%%%%%%%%%%%%%%%%%%%%%%%%%%%%%%%%%%%%%%%%%%%%%%%%%%%%%%%%%%%%%

\section{The $t^{(8)}$ tensor}
\label{app-t8}

The components Hodge dual of a Lorentz 2-form
$A=\tfrac{1}{2}A_{\mu\nu}dx^{\mu}\wedge dx^{\nu}$ are defined by 
\begin{equation}
\label{eq:Hodgedualdef}
\tilde{A}_{\mu\nu}\equiv
\tfrac{1}{2}\epsilon_{\mu\nu\rho\sigma}A^{\rho\sigma}\, ,
\hspace{1cm}
\tilde{\tilde{A}}=-A\, ,  
\end{equation}
and those of its self- and anti-self-dual parts are defined by
\begin{equation}
\label{eq:selfdualdef}
A^{\pm} = \tfrac{1}{2}(A\pm i\tilde{A})\, ,  
\hspace{1cm}
\tilde{A}^{\pm} =\mp iA^{\pm}\, .
\end{equation}

These definitions imply, for any pair of (real or complex) 2-forms $A$ and
$B$:
\begin{eqnarray}
\tilde{A}\tilde{B}
& = &
BA-\tfrac{1}{2}\textrm{Tr}_{L}(AB)\mathbbm{1}\, ,\\
\nonumber \\
\tilde{A} B 
& = & 
-\tilde{B}A+\tfrac{1}{2}\textrm{Tr}_{L}(A\tilde{B})\mathbbm{1}\, ,\\
 \nonumber \\
A^{\pm}B^{\pm} 
& = & 
-B^{\pm}A^{\pm}+\tfrac{1}{2}\textrm{Tr}_{L} (A^{\pm}B^{\pm})\mathbbm{1}\, ,\\
\nonumber \\
A^{\pm}B^{\mp}
& = & 
B^{\mp}A^{\pm}\, ,
\end{eqnarray}
where $AB$ stands for $A_{\mu}{}^{\nu}B_{\nu}{}^{\rho}$, $\mathbbm{1}$ stands
for $\delta_{\mu}{}^{\rho}$, $\textrm{Tr}_{L}(AB)$ stands for
$A_{\mu}{}^{\nu}B_{\nu}{}^{\mu}$ etc.

The $t^{(8)}$ tensor is a tensor totally symmetric in four pairs of
antisymmetric indices that can be defined by its contraction with 4 arbitrary
Lorentz 2-forms $A,B,C,D$:
\begin{eqnarray}
t^{(8)}{}_{abcd}A^{a}B^{b}C^{c}D^{d}
& = & 
8[\textrm{Tr}_{L}(ABCD)
+\textrm{Tr}_{L}(ACBD)
+\textrm{Tr}_{L}(ACDB)]
\nonumber\\
\nonumber \\
& &
-2[ \textrm{Tr}_{L}(AB)\textrm{Tr}_{L}(CD)
+\textrm{Tr}_{L}(AC)\textrm{Tr}_{L}(BD)
\nonumber \\
& & \nonumber \\
& & 
+\textrm{Tr}_{L}(AD)\textrm{Tr}_{L}(BC)]\, .
\end{eqnarray}
It is convenient to use only one Latin index $a,b,c\ldots$ to denote each of
these four pairs and write $t^{(8)}{}_{abcd}$ instead of
$t^{(8)}{}_{\mu_{1}\nu_{1}\mu_{2}\nu_{2}\mu_{3}\nu_{3}\mu_{4}\nu_{4}}=
t^{(8)}{}_{[\mu_{1}\nu_{1}][\mu_{2}\nu_{2}][\mu_{3}\nu_{3}][\mu_{4}\nu_{4}]}$.
Then, in terms of these indices, $t^{(8)}$ is completely symmetric
$t^{(8)}{}_{abcd}= t^{(8)}{}_{(abcd)}$.

Using the above relations plus the invariance of the trace under transposition
and the antisymmetry of the tensors $A,B,C,D$, one finds the identity
\begin{eqnarray}
\label{eq:t8identity}
t^{(8)}{}_{abcd}A^{a}B^{b}C^{c}D^{d}
& = &   
4\textrm{Tr}_{L}(A^{+}B^{+})\textrm{Tr}_{L}(C^{-}D^{-})
+4\textrm{Tr}_{L}(A^{+}C^{+})\textrm{Tr}_{L}(B^{-}D^{-})
\nonumber \\
& & \nonumber \\
& & 
+4\textrm{Tr}_{L}(A^{+}D^{+})\textrm{Tr}_{L}(B^{-}C^{-})
+(+\leftrightarrow -)\, .
\end{eqnarray}
That is: the only terms that contribute are those with zero helicity.  For the
particular choice $A=B=\partial_{\mu}F$, $C=D=\partial_{\nu}F$ the above
identity gives
\begin{eqnarray}
\label{eq:amplitude}
t^{(8)}{}_{abcd}\partial_{\mu}F^{a}\partial^{\mu}F^{b}
\partial_{\nu}F^{c}\partial^{\nu}F^{d}
& = &  
4\textrm{Tr}_{L}(\partial_{\mu}F^{+}\partial^{\mu}F^{+})
\textrm{Tr}_{L}(\partial_{\nu}F^{-}\partial^{\nu}F^{-})
\nonumber \\
& & \nonumber \\
& & 
+8\textrm{Tr}_{L}(\partial_{\mu}F^{+}\partial_{\nu}F^{+})
\textrm{Tr}_{L}(\partial^{\mu}F^{-}\partial^{\nu}F^{-})
+\mathrm{c.c}\, .
\end{eqnarray}
We also find
\begin{eqnarray}
t^{(8)}{}_{abcd}\partial_{\mu}F^{+a}\partial^{\mu}F^{-b}
\partial_{\nu}F^{+c}\partial^{\nu}F^{-d}
& = &  
8\textrm{Tr}_{L}(\partial_{\mu}F^{+}\partial_{\nu}F^{+})
\textrm{Tr}_{L}(\partial^{\mu}F^{-}\partial^{\nu}F^{-})
\, ,
\\
& & \nonumber \\
t^{(8)}{}_{abcd}\partial_{\mu}F^{+a}\partial^{\mu}F^{+b}
\partial_{\nu}F^{-c}\partial^{\nu}F^{-d}
& = &  
8\textrm{Tr}_{L}(\partial_{\mu}F^{+}\partial^{\mu}F^{+})
\textrm{Tr}_{L}(\partial_{\nu}F^{-}\partial^{\nu}F^{-})
\, ,
\end{eqnarray}
which, combined with the previous identity give
\begin{eqnarray}
t^{(8)}{}_{abcd}\partial_{\mu}F^{a}\partial^{\mu}F^{b}
\partial_{\nu}F^{c}\partial^{\nu}F^{d}
& = &  
2t^{(8)}{}_{abcd}\partial_{\mu}F^{+a}\partial^{\mu}F^{-b}
\partial_{\mu}F^{+c}\partial^{\nu}F^{-d}
\nonumber \\
& & \nonumber \\
& & 
+t^{(8)}{}_{abcd}\partial_{\mu}F^{+a}\partial^{\mu}F^{+b}
\partial_{\mu}F^{-c}\partial^{\nu}F^{-d}
\, ,
\end{eqnarray}
which are the initial sources of deformation used in \cite{Chemissany:2011yv}.

Finally, we can rewrite Eq.~(\ref{eq:amplitude}) up to total derivatives and up
to equations of motion in the form
\begin{eqnarray}
t^{(8)}{}_{abcd}\partial_{\mu}F^{a}\partial^{\mu}F^{b}
\partial_{\nu}F^{c}\partial^{\nu}F^{d}
& = &  
32\textrm{Tr}_{L}(F^{+}\partial_{\mu}\partial_{\nu}F^{+})
\textrm{Tr}_{L}(F^{-}\partial^{\mu}\partial^{\nu}F^{-})
\nonumber \\
& & \nonumber \\
& & 
+8[\textrm{Tr}_{L}(\partial_{\mu}F^{+}\partial_{\nu}F^{+})
\textrm{Tr}_{L}(F^{-}\partial^{\mu}\partial^{\nu}F^{-})
+\mathrm{c.c}]\, ,
\end{eqnarray}
which is basically the expression in \cite{Freedman:2011uc} reduced to $U(1)$.

%%%%%%%%%%%%%%%%%%%%%%%%%%%%%%%%%%%%%%%%%%%%%%%%%%%%%%%%%%%%%%%%%%%%%%
%%%%%%%%%%%%%%%%%%%%%%%%%%%%%%%%%%%%%%%%%%%%%%%%%%%%%%%%%%%%%%%%%%%%%%
%%%%%%%%%%%%%%%%%%%%%%%%%%%%%%%%%%%%%%%%%%%%%%%%%%%%%%%%%%%%%%%%%%%%%%
%%%%%%%%%%%%%%%%%%%%%%%%%%%%%%%%%%%%%%%%%%%%%%%%%%%%%%%%%%%%%%%%%%%%%%
%%%%%%%%%%%%%%%%%%%%%%%%%%%%%%%%%%%%%%%%%%%%%%%%%%%%%%%%%%%%%%%%%%%%%%

\section{$E_{7(7)}$\, and split octonions }
\label{ap-oct}

\vspace{.2cm}

\noindent
{\it A digression on commutators,  associators and triple products}

\vspace{.2cm}

A product (and the corresponding algebra) is {\bf commutative} if $xy=yx$ for
all $x, y$.  The {\bf commutator} is defined as
\begin{equation}
[x, y] \equiv xy-yx
\end{equation}
and it measures how far two elements are from commuting: $x$ and $y$ commute
iff their commutator is zero, $[x, y]=0$, the algebra is commutative iff all
commutators vanish. The role of the commutators in Physics is well known.

The product (and the algebra) is {\bf associative} if $(xy)z= x(yz)$ for all
$x,y,z$, in which case we drop all parentheses and write the product as
$xyz$. It is common to introduce the {\bf associator}
\begin{equation}
[x,y,z]\equiv (xy) z- x(yz)
\end{equation}
which measures how far three elements are from associating, $x,y,z$ associate
iff their associator is zero.

Some algebras are alternative, for example, the octonions algebra is only {\bf
  slightly non-associative} is {\bf alternative}: $[x_{1}, x_{2},x_{3}]=(-)^\pi
[x_{1\pi}, x_{2\pi},x_{3\pi}]$ where $\pi\in S_{3}$.  It means that whereas
the associator does not vanish, in general, it does vanish when two of its
variables are equal, for example
\begin{equation}
[x,y,x]= (xy) x- x(yx)=0\, ,  \qquad  [x,x,z]= (x^{2}) z- x(xz)=0
\end{equation}
The Jordan algebra has a particular associator vanishing (and it is sometimes
called {\bf power-associative} for it):
\begin{equation}
[x, y, x^{2}] = (x\circ y) \circ x^{2}- x\circ (y \circ x^{2})
\end{equation}
Here a symmetric product is defined as $x\circ y= y\circ x$ and $x^{2}= x\circ x$.

Finally, there is a Jordan triple product, defined as
\begin{equation}
\{ x, y, z\} = x \circ (y\circ z)+ z\circ (y\circ x) - (x\circ z) \circ y
\end{equation}

\vspace{.2cm}

\noindent
{\it Split octonions }

\vspace{.2cm}

For the split octonions, which are relevant to $E_{7(7)}$, one defines the
general element
\begin{equation}
O^{s}
= 
o_{0}
+ 
o_{1} j_{1}
+
o_{2} j_{2}
+ 
o_{3} j_{3}
+ 
o_{4} j_{4}^{s}
+ 
o_{5} j_{5}^{s}
+ 
o_{6} j_{6}^{s}
+ o_{7} j_{7}^{s}\,,
\end{equation}
and 
\begin{equation}
\overline{O}^{\, s}
= 
o_{0}
- 
o_{1} j_{1}
-
o_{2} j_{2}
- 
o_{3} j_{3}
- 
o_{4} j_{4}^{s}
- 
o_{5} j_{5}^{s}
- 
o_{6} j_{6}^{s}
- 
o_{7} j_{7}^{s}\, ,
\end{equation}
where the rules of multiplication of the seven ``imaginary'' units are 
\begin{equation}
j_{l} \, j_{k} 
= 
-\delta_{kl}+ \eta_{klm} j_{m}\, ,
\hspace{.5cm}
j_{\mu}^{s} \, j_{\nu}^{s} = \delta_{\mu\nu}- \eta_{\mu\nu m} j_{m}\, ,
\hspace{.5cm} 
j_{l} \, j_{\mu}^{s} =  \eta_{m\mu\nu} j_{\nu}^{s}\, ,
\end{equation}
where $l, k, m=1,2,3$, $ \mu, \nu =4,5,6,7$ and $\eta_{ABC}$, with
$A,B,C=1,\cdots,7$, is completely antisymmetric and given by
\begin{equation}
\eta_{ABC}
= 
1\,\,\, \mathrm{when}\,\,\, (ABC)= (123), (471), (572), (673), (624), (435),
(516)\, .
\end{equation}
In particular
\begin{equation}
j_{1}^{2}
=
j_{2}^{2}
=
j_{3}^{2}
=
-1\, ,
\hspace{.5cm} 
(j_{4}^{s})^{2}
=
(j_{5}^{s})^{2}
=
(j_{6}^{s})^{2}
=
(j_{7}^{s})^{2}
=
1\, .
\end{equation}

For real octonions

\begin{equation}
O
= 
o_{0}
+ 
o_{1} j_{1}
+
o_{2} j_{2}
+ 
o_{3} j_{3}
+ 
o_{4} j_{4}
+ 
o_{5} j_{5}
+ 
o_{6} j_{6}
+ 
o_{7} j_{7}\,. ,
\end{equation}
and
\begin{equation}
\overline{O}
= 
o_{0}
- 
o_{1} j_{1}
-
o_{2} j_{2}
- 
o_{3} j_{3}
- 
o_{4} j_{4}
- 
o_{5} j_{5}
- 
o_{6} j_{6}
- 
o_{7} j_{7}\, ,
\end{equation}
all the imaginary units square to minus one
\begin{equation}
j_{A}^{2}=-1\, ,\,\,\, \forall A= 1, \cdots , 7\, . 
\end{equation}
Therefore, for real octonions
\begin{equation}
O\overline{O} = \sum_{A=0}^{7} O_{A}^{2} 
\end{equation}
whereas for the split ones
\begin{equation}
 O^{s}\overline{O}^{\, s}
= 
\sum_{B=0}^{3} O_{B}^{2}- \sum_{C=4}^{7} O_{C}^{2}\, .
\end{equation}

\vspace{.2cm}

\noindent
{\it Freudenthal/Jordan basis}

\vspace{.2cm}

The fundamental ${\bf 56}$ of $E_{7(7)}$\, can be repackaged using the
Freudenthal triple systems and split octonions, as follows.  From 28 vectors
and 28 dual vectors we separate the zero direction from the other 27, in
agreement with the decomposition of $E_{7(7)}$\, via $E_{6(6)}$. The
decomposition involves the $\mathfrak{so}(1,1)$ grading
\begin{equation}
\label{e77}
\mathfrak{e}_{7,7}= \mathfrak{e}_{6,6}+\mathfrak{so}(1,1)+ \mathfrak{p}\, ,
\hspace{.5cm}
\mathfrak{p}=\mathbf{27_{-2}}+\mathbf{27^{\prime}_{+2}}\, ,
\end{equation}
and $\mathfrak{p}$ carries the representations of
$\mathfrak{e}_{6,6}+\mathfrak{so}(1,1)$. The $\mathbf{56}$ of $E_{7(7)}$\,
takes the form
\begin{eqnarray}
\label{E66}
X
= 
\left(
\begin{array}{cc}
F^{0} & F^{I} \\
G^{I} & G^{0} \\
\end{array}
\right) 
\Rightarrow 
\left(
\begin{array}{cc}
\alpha     & \mathbf{x} \\
\mathbf{y} &   \beta    \\
\end{array}
\right)\, .
\end{eqnarray}
 
Here the $3 \times 3$ split octonionic Hermitian matrices $\mathbf{x}$ and
$\mathbf{y}$ are elements of the the split exceptional Jordan algebra of
degree three, $J_{3}^{\mathbb{O}_s}$.  The 27 objects in $\mathbf{x}$ and 27
in $\mathbf{y}$ are organized in terms of 3 split octonions ($3\times 8=24$)
and 3 numbers, each, so that
\begin{eqnarray}
\label{x}
\mathbf{x}
= 
\left(
\begin{array}{ccc}
\alpha_{1} & O_{1}^{s} & \overline{O}_{2}^{\, s} \\
\overline{O}_{1}^{\, s} & \beta_{1} & O_{3}^{s} \\
O_{2}^{s} & \overline{O}_{3}^{\,s} & \gamma_{1} \\
\end{array}
\right)\, ,
\end{eqnarray}
and there is an analogous expression for $\mathbf{y}$.

The four-linear form derived in \cite{Faulkner} is given by the symplectic
product of the triple product with the fourth element
\begin{equation}
J^{\prime}_{4}[X_{1}, X_{2}, X_{3}, X_{4}]
= 
\left\langle \left( X_{1}, X_{2}, X_{3} \right) ,X_{4}\right\rangle\, ,
\end{equation}
where the definition of the Freudenthal triple, $(X_{1}, X_{2}, X_{3})$, a ternary
product, corresponds to multiplication of the 3 elements, taken in the form
(\ref{E66}), (\ref{x}).

\vspace{.2cm}

\noindent
{\it Fano basis}

\vspace{.2cm}

The fundamental ${\bf 56}$ of $E_{7(7)}$\, can be repackaged using the
decomposition $E_{7(7)}\supset \Big (SL(2, \mathbb{R})\Big)^{7}$,
\cite{Duff:2006ue}. In such case there is a relation between Alice, Bob,
Charlie, Daisy, Emma, Fred and George and 7 qubits in quantum information
theory. This is also related to the fact that the fundamental ${\bf 56}$ of
$E_{7(7)}$\, can be repackaged into 7 subsets with 8 elements each, as shown
in \cite{Borsten:2008wd}, \cite{Levay:2008mi}. For example, in the notation of
\cite{Levay:2008mi} with
\begin{eqnarray}
A
& = &
({\bf a}_{0}, {\bf a}_{1}, {\bf a}_{2}, {\bf a}_{3}, {\bf a}_{4}, {\bf a}_{5},
{\bf a}_{6}, {\bf a}_{7})\, ,
\\
\vdots &  \vdots &  \hspace{2cm} \vdots \nonumber \\
G
& = &
({\bf g}_{0}, {\bf g}_{1}, {\bf g}_{2}, {\bf g}_{3}, {\bf g}_{4}, {\bf g}_{5},
{\bf g}_{6}, {\bf g}_{7})\, ,
\end{eqnarray}
we can write
\begin{eqnarray}
F^{ij}
& = &
\begin{pmatrix}
0&-{\bf a}_{7}&-{\bf b}_{7}&-{\bf c}_{7}&-{\bf d}_{7}&-{\bf e}_{7}&-{\bf f}_{7}&-{\bf g}_{7}\\
{\bf a}_{7}&0&{\bf f}_{1}&{\bf d}_{4}&-{\bf c}_{2}&{\bf g}_{2}&-{\bf b}_{4}&-{\bf e}_{1}\\
{\bf b}_{7}&-{\bf f}_{1}&0&{\bf g}_{1}&{\bf e}_{4}&-{\bf d}_{2}&{\bf a}_{2}&-{\bf c}_{4}\\
{\bf c}_{7}&-{\bf d}_{4}&-{\bf g}_{1}&0&{\bf a}_{1}&{\bf f}_{4}&-{\bf e}_{2}&{\bf b}_{2}\\
{\bf d}_{7}&{\bf c}_{2}&-{\bf e}_{4}&-{\bf a}_{1}&0&{\bf b}_{1}&{\bf g}_{4}&-{\bf f}_{2}\\
{\bf e}_{7}&-{\bf g}_{2}&{\bf d}_{2}&-{\bf f}_{4}&-{\bf b}_{1}&0&{\bf c}_{1}&{\bf a}_{4}\\
{\bf f}_{7}&{\bf b}_{4}&-{\bf a}_{2}&{\bf e}_{2}&-{\bf g}_{4}&-{\bf c}_{1}&0&{\bf d}_{1}\\
{\bf g}_{7}&{\bf e}_{1}&{\bf c}_{4}&-{\bf b}_{2}&{\bf f}_{2}&-{\bf a}_{4}&-{\bf d}_{1}&0
\end{pmatrix}\, ,
%\label{x}
\\
& & \nonumber \\
G_{ij}
& = &
\begin{pmatrix}
0&-{\bf a}_{0}&-{\bf b}_{0}&-{\bf c}_{0}&-{\bf d}_{0}&-{\bf e}_{0}&-{\bf f}_{0}&-{\bf g}_{0}\\
{\bf a}_{0}&0&{\bf f}_{6}&{\bf d}_{3}&-{\bf c}_{5}&{\bf g}_{5}&-{\bf b}_{3}&-{\bf e}_{6}\\
{\bf b}_{0}&-{\bf f}_{6}&0&{\bf g}_{6}&{\bf e}_{3}&-{\bf d}_{5}&{\bf a}_{5}&-{\bf c}_{3}\\
{\bf c}_{0}&-{\bf d}_{3}&-{\bf g}_{6}&0&{\bf a}_{6}&{\bf f}_{3}&-{\bf e}_{5}&{\bf b}_{5}\\
{\bf d}_{0}&{\bf c}_{5}&-{\bf e}_{3}&-{\bf a}_{6}&0&{\bf b}_{6}&{\bf g}_{3}&-{\bf f}_{5}\\
{\bf e}_{0}&-{\bf g}_{5}&{\bf d}_{5}&-{\bf f}_{3}&-{\bf b}_{6}&0&{\bf c}_{6}&{\bf a}_{3}\\
{\bf f}_{0}&{\bf b}_{3}&-{\bf a}_{5}&{\bf e}_{5}&-{\bf g}_{3}&-{\bf c}_{6}&0&{\bf d}_{6}\\
{\bf g}_{0}&{\bf e}_{6}&{\bf c}_{3}&-{\bf b}_{5}&{\bf f}_{5}&-{\bf a}_{3}&-{\bf d}_{6}&0
\end{pmatrix}\, .
\label{y}
\end{eqnarray}
This form of the $E_{7(7)}$\, doublet was derived from the Coxeter
sub-geometry of the split Cayley hexagon and using the 7-fold symmetry of the
Coxeter graph.  In this form our 4-linear $E_{7(7)}$\, invariant
(\ref{eq:J4prime}) consists of a set of products of 4 terms where each of the
$F$ and $G$ is represented by $8\times 8$ matrices above.

%%%%%%%%%%%%%%%%%%%%%%%%%%%%%%%%%%%%%%%%%%%%%%%%%%%%%%%%%%%%%%%%%%%%%%
%%%%%%%%%%%%%%%%%%%%%%%%%%%%%%%%%%%%%%%%%%%%%%%%%%%%%%%%%%%%%%%%%%%%%%
%%%%%%%%%%%%%%%%%%%%%%%%%%%%%%%%%%%%%%%%%%%%%%%%%%%%%%%%%%%%%%%%%%%%%%
%%%%%%%%%%%%%%%%%%%%%%%%%%%%%%%%%%%%%%%%%%%%%%%%%%%%%%%%%%%%%%%%%%%%%%
%%%%%%%%%%%%%%%%%%%%%%%%%%%%%%%%%%%%%%%%%%%%%%%%%%%%%%%%%%%%%%%%%%%%%%
%%%%%%%%%%%%%%%%%%%%%%%%%%%%%%%%%%%%%%%%%%%%%%%%%%%%%%%%%%%%%%%%%%%%%%
%%%%%%%%%%%%%%%%%%%%%%%%%%%%%%%%%%%%%%%%%%%%%%%%%%%%%%%%%%%%%%%%%%%%%%
%%%%%%%%%%%%%%%%%%%%%%%%%%%%%%%%%%%%%%%%%%%%%%%%%%%%%%%%%%%%%%%%%%%%%%
%%%%%%%%%%%%%%%%%%%%%%%%%%%%%%%%%%%%%%%%%%%%%%%%%%%%%%%%%%%%%%%%%%%%%%
%%%%%%%%%%%%%%%%%%%%%%%%%%%%%%%%%%%%%%%%%%%%%%%%%%%%%%%%%%%%%%%%%%%%%%

%%%%%%%%%%%%%%%%%%%%%%%%%%%%%%%%%%%%%%%%%%%%%%%%%%%%%%%%%%%%%%
%
%

\begin{thebibliography}{99}
\raggedright

%\cite{Bern:2007hh} 
\bibitem{Bern:2007hh} 
Z.~Bern, J.~J.~Carrasco, L.~J.~Dixon, H.~Johansson, D.~A.~Kosower and R.~Roiban,
``Three-Loop Superfiniteness of $\mathcal{N}=8$ Supergravity,''
Phys.\ Rev.\ Lett.\  {\bf 98}, 161303 (2007)
[\hepth{0702112}].
%%CITATION = HEP-TH/0702112;%%

%\cite{Bern:2008pv}
\bibitem{Bern:2008pv} 
Z.~Bern, J.~J.~M.~Carrasco, L.~J.~Dixon, H.~Johansson and R.~Roiban,
``Manifest Ultraviolet Behavior for the Three-Loop Four-Point Amplitude of $\mathcal{N}=8$ Supergravity,''
Phys.\ Rev.\ D {\bf 78}, 105019 (2008)
[\arxiv{0808.4112} [hep-th]].
%%CITATION = ARXIV:0808.4112;%%

%\cite{Bern:2012cd}
\bibitem{Bern:2012cd}
Z.~Bern, S.~Davies, T.~Dennen and Y.~-t.~Huang,
``Absence of Three-Loop Four-Point Divergences in $\mathcal{N}=4$ Supergravity,''
\arxiv{1202.3423} [hep-th].
%%CITATION = ARXIV:1202.3423;%%
                     
%\cite{Tourkine:2012ip}
\bibitem{Tourkine:2012ip}
P.~Tourkine and P.~Vanhove,
``A $R^4$ non-renormalisation theorem in N=4 supergravity,''
\arxiv{1202.3692} [hep-th].
%%CITATION = ARXIV:1202.3692;%%      
  
%\cite{Cremmer:1979up}
\bibitem{Cremmer:1979up}
 E.~Cremmer and B.~Julia,
``The SO(8) Supergravity,''
Nucl.\ Phys.\  B {\bf 159}, 141 (1979).
%%CITATION = NUPHA,B159,141;%%
  
%\cite{de Wit:1982ig}
\bibitem{de Wit:1982ig}
B.~de Wit and H.~Nicolai,
``$\mathcal{N}=8$ Supergravity,''
Nucl.\ Phys.\  B {\bf 208}, 323 (1982).
%%CITATION = NUPHA,B208,323;%%
 
%\cite{Gaillard:1981rj}
\bibitem{Gaillard:1981rj}
M.~K.~Gaillard and B.~Zumino, 
`` Duality Rotations For Interacting Fields,''
Nucl.\ Phys.\  B {\bf 193}, 221 (1981).
%%CITATION = NUPHA, B193, 221;%%
P.~Aschieri, S.~Ferrara and B.~Zumino, 
`` Duality Rotations in Nonlinear Electrodynamics and in Extended
Supergravity,''
Riv.\ Nuovo Cim.\  {\bf 31}, 625 (2008)
[\arxiv{0807.4039}] [hep-th]].
%%CITATION = RNCIB, 31, 625;%%

%\cite{Kallosh:2011dp}
\bibitem{Kallosh:2011dp} 
  R.~Kallosh,
  ``$E_{7(7)}$ Symmetry and Finiteness of $\mathcal{N}=8$ Supergravity,''
  JHEP {\bf 1203}, 083 (2012)
  [\arxiv{1103.4115}] [hep-th].
%%CITATION = ARXIV:1103.4115;%%
R.~Kallosh, 
``${\cal N}=8$ Counterterms and $E_{7(7)}$ Current Conservation,''
JHEP {\bf 1106}, 073 (2011)
[\arxiv{1104.5480}] [hep-th]].
%%CITATION = JHEPA, 1106, 073;%%
 
%\cite{Bossard:2011ij}
\bibitem{Bossard:2011ij}
G.~Bossard and H.~Nicolai,
``Counterterms vs. Dualities,''
JHEP {\bf 1108} (2011) 074
[\arxiv{1105.1273} [hep-th]].
%%CITATION = ARXIV:1105.1273;%%

%\cite{Carrasco:2011jv}
\bibitem{Carrasco:2011jv} 
J.~J.~M.~Carrasco, R.~Kallosh and R.~Roiban,
``Covariant procedures for perturbative non-linear deformation of duality-invariant theories,''
Phys.\ Rev.\ D {\bf 85}, 025007 (2012)
[\arxiv{1108.4390} [hep-th]].
%%CITATION = ARXIV:1108.4390;%%


%\cite{Chemissany:2011yv}
\bibitem{Chemissany:2011yv} 
W.~Chemissany, R.~Kallosh and T.~Ort\'{\i}n,
``Born-Infeld with Higher Derivatives,''
Phys.\ Rev.\ D {\bf 85}, 046002 (2012)
[\arxiv{1112.0332} [hep-th]].
%%CITATION = ARXIV:1112.0332;%%
  
%\cite{Broedel:2012gf}
\bibitem{Broedel:2012gf} 
J.~Broedel, J.~J.~M.~Carrasco, S.~Ferrara, R.~Kallosh and R.~Roiban,
``$\mathcal{N}=2$ Supersymmetry and U(1)-Duality,''
\arxiv{1202.0014} [hep-th].
%%CITATION = ARXIV:1202.0014;%%
  
\bibitem{Kallosh:2008ru}
R.~Kallosh, C.~H.~Lee and T.~Rube,
``$\mathcal{N}=8$ Supergravity 4-point Amplitudes,''
JHEP {\bf 0902}, 050 (2009)
[\arxiv{0811.3417} [hep-th]].
%%CITATION = JHEPA,0902,050;%%

%\cite{Freedman:2011uc}
\bibitem{Freedman:2011uc} 
D.~Z.~Freedman and E.~Tonni,
 ``The $D^{2k} R^{4}$ Invariants of $\mathcal{N}=8$ Supergravity,''
JHEP {\bf 1104}, 006 (2011)
[\arxiv{1101.1672} [hep-th]].
%%CITATION = ARXIV:1101.1672;%%
  
%\cite{Brink:1979nt}
\bibitem{Brink:1979nt}
L.~Brink and P.~S.~Howe,
``The ${\cal{N}}=8$ Supergravity In Superspace,''
Phys.\ Lett.\  B {\bf 88}, 268 (1979).
%%CITATION = PHLTA,B88,268;%%
  
%\cite{Howe:1981gz}
\bibitem{Howe:1981gz} 
P.~S.~Howe,
``Supergravity In Superspace,''
Nucl.\ Phys.\ B {\bf 199}, 309 (1982).
%%CITATION = NUPHA,B199,309;%%


\bibitem{Brown} R.B.~Brown, ``Groups of type $E7$'', Journal fur die reine und
  angewandte Mathematik, {\bf 236}, 79 (1969).

%\cite{Ferrara:2011dz}
\bibitem{Ferrara:2011dz} 
S.~Ferrara and R.~Kallosh,
``Creation of Matter in the Universe and Groups of Type $E7$,''
JHEP {\bf 1112}, 096 (2011)
[\arxiv{1110.4048} [hep-th]].
%\cite{Ferrara:2011aa}
%\bibitem{Ferrara:2011aa} 
  S.~Ferrara and A.~Marrani,
  ``Black Holes and Groups of Type $E_7$,''
  \arxiv{1112.2664} [hep-th].
  %%CITATION = ARXIV:1112.2664;%%
%%CITATION = ARXIV:1110.4048;%%    
S.~Ferrara, R.~Kallosh and A.~Marrani,
``Degeneration of Groups of Type $E7$ and Minimal Coupling in Supergravity,''
\arxiv{1202.1290} [hep-th].
%%CITATION = ARXIV:1202.1290;%%

%\cite{Kallosh:2012ei}
\bibitem{Kallosh:2012ei} 
R.~Kallosh,
``On Absence of 3-loop Divergence in $\mathcal{N}=4$ Supergravity,''
\arxiv{1202.4690} [hep-th].
%%CITATION = ARXIV:1202.4690;%%

\bibitem{CARTAN}
E.~Cartan,
OEuvres compl`etes,
Paris: Editions du Centre National de la Recherche Scientifique, 1984.

%\cite{Kallosh:1996uy}
\bibitem{Kallosh:1996uy} 
R.~Kallosh and B.~Kol,
``E(7) symmetric area of the black hole horizon,''
Phys.\ Rev.\ D {\bf 53}, 5344 (1996)
[\hepth{9602014}].
%%CITATION = HEP-TH/9602014;%%
M.~Cvetic and C.~M.~Hull,
``Black holes and U duality,''
Nucl.\ Phys.\ B {\bf 480}, 296 (1996)
[\hepth{9606193}].
%%CITATION = HEP-TH/9606193;%%

%\cite{Bianchi:2009wj}
\bibitem{Bianchi:2009wj} 
M.~Bianchi, S.~Ferrara and R.~Kallosh,
 ``Perturbative and Non-perturbative N =8 Supergravity,''
Phys.\ Lett.\ B {\bf 690}, 328 (2010)
[\arxiv{0910.3674} [hep-th]].
%%CITATION = ARXIV:0910.3674;%%

%\cite{Gunaydin:2000xr}
\bibitem{Gunaydin:2000xr}
M.~G\"unaydin, K.~Koepsell and H.~Nicolai,
``Conformal and quasiconformal realizations of exceptional Lie groups,''
Commun.\ Math.\ Phys.\  {\bf 221}, 57 (2001)
[\hepth{0008063}].
%%CITATION = HEP-TH/0008063;%%   

%\cite{Gunaydin:2009zza}
\bibitem{Gunaydin:2009zza} 
M.~G\"unaydin and O.~Pavlyk,
``Quasiconformal Realizations of E(6)(6), E(7)(7), E(8)(8) and SO(n+3,m+3), $\mathcal{N}=4$ Supergravity and Spherical Vectors,''
\arxiv{0904.0784} [hep-th].
%%CITATION = ARXIV:0904.0784;%%

%\cite{Kallosh:1980fi}
\bibitem{Kallosh:1980fi}
R.~E.~Kallosh,
``Counterterms in extended supergravities,''
Phys.\ Lett.\  B {\bf 99} (1981) 122;
%%CITATION = PHLTA,B99,122;%%
P.~S.~Howe and U.~Lindstrom,
``Higher Order Invariants In Extended Supergravity,''
Nucl.\ Phys.\  B {\bf 181}, 487 (1981).
%%CITATION = NUPHA,B181,487;%%
P.~S.~Howe, K.~S.~Stelle and P.~K.~Townsend,
``Superactions,''
Nucl.\ Phys.\  B {\bf 191}, 445 (1981).
%%CITATION = NUPHA,B191,445;%%


\bibitem{Shmakova:1999ai}
M.~Shmakova,
``One loop corrections to the D3-brane action,''
Phys.\ Rev.\ D {\bf 62}, 104009 (2000)
[\hepth{9906239}].
%%CITATION = HEP-TH/9906239;%%
  
\bibitem{DeGiovanni:1999hr} 
A.~De Giovanni, A.~Santambrogio and D.~Zanon,
``$\alpha'{}^{4}$ corrections to the $\mathcal{N}=2$ supersymmetric Born-Infeld action,''
Phys.\ Lett.\ B {\bf 472}, 94 (2000)
[Erratum-ibid.\ B {\bf 478}, 457 (2000)]
[\hepth{9907214}].
%%CITATION = HEP-TH/9907214;%%
  
%\cite{Kallosh:2008ic}
\bibitem{Kallosh:2008ic}
R.~Kallosh and M.~Soroush,
``Explicit Action of E7(7) on $\mathcal{N}=8$ Supergravity Fields,''
Nucl.\ Phys.\  B {\bf 801}, 25 (2008)
[\arxiv{0802.4106} [hep-th]].
%%CITATION = NUPHA,B801,25;%%

\bibitem{Beisert:2010jx}
N.~Beisert, H.~Elvang, D.~Z.~Freedman, M.~Kiermaier, A.~Morales, S.~Stieberger,
``E7(7) constraints on counterterms in $\mathcal{N}=8$ supergravity,''
Phys.\ Lett.\  {\bf B694}, 265-271 (2010).
[\arxiv{1009.1643} [hep-th]].
H.~Elvang, D.~Z.~Freedman and M.~Kiermaier,
``A simple approach to counterterms in $\mathcal{N}=8$ supergravity,''
JHEP {\bf 1011}, 016 (2010)
[\arxiv{1003.5018} [hep-th]].
%%CITATION = JHEPA,1011,016;%%

%\cite{Meessen:2010fh}
\bibitem{Meessen:2010fh}
P.~Meessen, T.~Ort\'{\i}n and S.~Vaula,
``All the timelike supersymmetric solutions of all ungauged d=4 supergravities,''
JHEP {\bf 1011} (2010) 072
[\arxiv{1006.0239} [hep-th]].
%%CITATION = ARXIV:1006.0239;%%

%\cite{Andrianopoli:1996ve}
\bibitem{Andrianopoli:1996ve}
L.~Andrianopoli, R.~D'Auria and S.~Ferrara,
``U duality and central charges in various dimensions revisited,''
Int.\ J.\ Mod.\ Phys.\ A {\bf 13} (1998) 431
[\hepth{9612105}].
%%CITATION = HEP-TH/9612105;%%
L.~Andrianopoli, R.~D'Auria and S.~Ferrara,
``U invariants, black hole entropy and fixed scalars,''
Phys.\ Lett.\ B {\bf 403}, 12 (1997)
[\hepth{9703156}].
%%CITATION = HEP-TH/9703156;%%

%\cite{Balasubramanian:1997az}
\bibitem{Balasubramanian:1997az} 
V.~Balasubramanian, F.~Larsen and R.~G.~Leigh,
``Branes at angles and black holes,''
Phys.\ Rev.\ D {\bf 57}, 3509 (1998)
\hepth{9704143}.
%%CITATION = HEP-TH/9704143;%%

%\cite{Ferrara:1997tw}
\bibitem{Ferrara:1997tw}
S.~Ferrara, G.~W.~Gibbons, R.~Kallosh,
``Black holes and critical points in moduli space,''
Nucl.\ Phys.\  {\bf B500 } (1997)  75-93.
[\hepth{9702103}].

%\cite{Ferrara:2006em}
\bibitem{Ferrara:2006em} 
S.~Ferrara and R.~Kallosh,
``On $\mathcal{N}=8$ attractors,''
Phys.\ Rev.\ D {\bf 73}, 125005 (2006)
[\hepth{0603247}].
%%CITATION = HEP-TH/0603247;%%

\bibitem {Faulkner} 
J.~Faulkner, 
`` A construction of Lie algebras from a class of ternary algebras.'' 
Trans.\ Am.\ Math.\ Soc., {\bf 155}, 397 (1971)

%\cite{Borsten:2008wd}
\bibitem{Borsten:2008wd} 
L.~Borsten, D.~Dahanayake, M.~J.~Duff, H.~Ebrahim and W.~Rubens,
``Black Holes, Qubits and Octonions,''
Phys.\ Rept.\  {\bf 471}, 113 (2009)
[\arxiv{0809.4685} [hep-th]].
%%CITATION = ARXIV:0809.4685;%%


%\cite{Andrianopoli:2011gy}
\bibitem{Andrianopoli:2011gy} 
L.~Andrianopoli, R.~D'Auria, S.~Ferrara, A.~Marrani and M.~Trigiante,
``Two-Centered Magical Charge Orbits,''
JHEP {\bf 1104}, 041 (2011)
[\arxiv{1101.3496} [hep-th]].
%%CITATION = ARXIV:1101.3496;%%

%\cite{Bellorin:2006xr}
\bibitem{Bellorin:2006xr}
J.~Bellor\'{\i}n, P.~Meessen, T.~Ort\'{\i}n
``Supersymmetry, attractors and cosmic censorship,''
Nucl.\ Phys.\  {\bf B762 } (2007)  229-255.
[\hepth{0606201}].           
%%CITATION = NUPHA,B762,229;%%

%\cite{Roiban:2012gi}
\bibitem{Roiban:2012gi} 
R.~Roiban and A.~A.~Tseytlin,
``On duality symmetry in perturbative quantum theory,''
[\arxiv{1205.0176} [hep-th]].
 %%CITATION = ARXIV:1205.0176;%%

\bibitem{Padua} P. Pasti,  D. Sorokin, and M. Tonin ``Covariant actions for models with nonÐlinear twisted selfÐduality'', to appear soon. 

%\cite{Cremmer:1980ru}
\bibitem{Cremmer:1980ru} 
E.~Cremmer and S.~Ferrara,
``Formulation of Eleven-Dimensional Supergravity in Superspace,''
Phys.\ Lett.\ B {\bf 91}, 61 (1980).
%%CITATION = PHLTA,B91,61;%%

%\cite{Duff:2006ue}
\bibitem{Duff:2006ue} 
M.~J.~Duff and S.~Ferrara,
``E(7) and the tripartite entanglement of seven qubits,''
Phys.\ Rev.\ D {\bf 76}, 025018 (2007)
[quant-ph 0609227].
%%CITATION = QUANT-PH/0609227;%%

%\cite{Levay:2008mi}
\bibitem{Levay:2008mi} 
P.~Levay, M.~Saniga and P.~Vrana,
``Three-Qubit Operators, the Split Cayley Hexagon of Order Two and Black Holes,''
Phys.\ Rev.\ D {\bf 78}, 124022 (2008)
[\arxiv{0808.3849} [quant-ph]].
%%CITATION = ARXIV:0808.3849;%% 

%%%%%%%%%%%%%%%%%%%%%%%%%%%%%%%%%%%%%%%%%%%%%%%%%%%%%%%%%%%%%%%%%%%%%%


%\cite{Green:1981xx}
\bibitem{Green:1981xx} 
M.~B.~Green and J.~H.~Schwarz,
``Supersymmetrical Dual String Theory. 2. Vertices and Trees,''
Nucl.\ Phys.\ B {\bf 198}, 252 (1982).
%%CITATION = NUPHA,B198,252;%%
J.~H.~Schwarz,
``Superstring Theory,''
Phys.\ Rept.\ \ {\bf 89}, 223  (1982).
%%CITATION = PRPLC,89,223;%%













%%%%%%%%%%%%%%%%%%%%%%%%%%%%%%%%%%%%%%%%%%%%%%%%%%%%%%%%%%%%%%%%%%%%%%


  















%%%%%%%%%%%%%%%%%%%%%%%%%%%%%%%%%%%%%%%%%%%%%%%%%%%%%%%%%%%%%%%%%%%%%%












  
  

  




% %\cite{Bianchi:2008pu}
% \bibitem{Bianchi:2008pu}
% M.~Bianchi, H.~Elvang and D.~Z.~Freedman,
% ``Generating Tree Amplitudes in $\mathcal{N}=4$ SYM and N = 8 SG,''
% JHEP {\bf 0809}, 063 (2008)
% [\arxiv{0805.0757} [hep-th]].
%   %%CITATION = JHEPA,0809,063;%%
% N.~Arkani-Hamed, F.~Cachazo and J.~Kaplan,
% ``What is the Simplest Quantum Field Theory?,''
% JHEP {\bf 1009}, 016 (2010)
% [\arxiv{0808.1446} [hep-th]].
% %%CITATION = JHEPA,1009,016;%%
% R.~Kallosh and T.~Kugo,
% ``The footprint of $E7$ in amplitudes of $\mathcal{N}=8$ supergravity,''
% JHEP {\bf 0901}, 072 (2009)
% [\arxiv{0811.3414} [hep-th]].
% %%CITATION = JHEPA,0901,072;%%

% %\cite{Brodel:2009hu}
% \bibitem{Brodel:2009hu}
% J.~Broedel and L.~J.~Dixon,
% ``$R^{4}$ counterterm and E7(7) symmetry in maximal supergravity,''
% JHEP {\bf 1005}, 003 (2010)
% [\arxiv{0911.5704} [hep-th]];
% %%CITATION = JHEPA,1005,003;%%
% H.~Elvang and M.~Kiermaier,
% ``Stringy KLT relations, global symmetries, and $E_{7(7)}$ violation,''
% \arxiv{1007.4813} [hep-th].
% %%CITATION = ARXIV:1007.4813;%%
% G.~Bossard, P.~S.~Howe and K.~S.~Stelle,
% ``On duality symmetries of supergravity invariants,''
% JHEP {\bf 1101}, 020 (2011)
% [\arxiv{1009.0743} [hep-th]].
% %%CITATION = JHEPA,1101,020;%%



  


  



%%%%%%%%%%%%%%%%%%%%%%%%%%%%%%%%%%%%%%%%%%%%%%%%%%%%%%%%%%%%%%
\end{thebibliography}
\end{document}